\documentclass[11pt,a4paper]{article}

\usepackage[margin=2.1cm]{geometry}
\usepackage{amsmath,amssymb,mathtools}
\usepackage{graphicx}
\usepackage{booktabs}
\usepackage{tabularx}
\usepackage{array}
\usepackage{multirow}
\usepackage{siunitx}
\usepackage{microtype}
\usepackage{hyperref}
\usepackage{xcolor}
\usepackage{enumitem}
\usepackage{float}
\usepackage{tikz}
\usetikzlibrary{arrows.meta,positioning,fit,calc}

\hypersetup{
  colorlinks=true,
  linkcolor=blue,
  citecolor=blue,
  urlcolor=blue,
  hypertexnames=false,
  pdftitle={Beyond RMSE and Granger Gain: Multi-Horizon Forecast-Error Entropy for Nonlinear Predictive Coupling},
  pdfauthor={Andrei Velichko}
}
\setlist{nosep}
\newcommand{\FEDE}{\mathrm{FEDE}}
\newcommand{\RMSE}{\mathrm{RMSE}}
\newcommand{\gmin}{\gamma_{\min}}

\title{Beyond RMSE and Granger Gain: Multi-Horizon Forecast-Error Entropy for Nonlinear Predictive Coupling}
\author{Andrei Velichko\\
\small Institute of Physics and Technology, Petrozavodsk State University, Petrozavodsk, Russia\\
\small Email: \href{mailto:velichkogf@gmail.com}{velichkogf@gmail.com} \quad
\href{https://scholar.google.com/citations?user=b4jL6-oAAAAJ}{Google Scholar profile}}
\date{Preprint, October 2026}

\begin{document}
\maketitle

\begin{abstract}
We investigate whether the past of an auxiliary time series provides predictive information about a target beyond that contained in the target's own history. We compare matched out-of-sample target-only and target-plus-source forecasts and introduce the forecast-error distribution entropy (FEDE) predictive-causality framework, where predictive causality is understood in the Granger sense: auxiliary history is informative when it improves prediction beyond the target's own past. FEDE quantifies how the full signed forecast-error distribution changes across prediction horizons. Controlled nonlinear benchmarks hide source--target dependence behind nonlinear transformations not supplied to the predictors. LightGBM recovers these hidden channels, whereas Ridge does not. Across four equal-power nonlinearities and four noise levels, the mean response-sensitivity statistic for FEDE remains 29.00, 27.47, 23.22, and 28.67, compared with 14.09, 10.68, 12.67, and 12.71 for mean log-RMSE gain. Directional and source controls separate the true source-to-target channel from the reverse direction and from independent periodic, quasiperiodic, chaotic, and autoregressive auxiliaries. Chronological Granger-style comparisons show strong model-form dependence: a degree-2 polynomial model is exceptionally sensitive to the matched quadratic channel but becomes weak or oppositely aligned for other nonlinearities, while basic convergent cross mapping is weak or directionally inconsistent in this stochastic, partly non-invertible setting. FEDE remains positively aligned across all 108 tested combinations of window length, history length, and maximum forecast horizon. Together, these results establish FEDE as a sensitive and robust horizon-resolved tool that extracts predictive information beyond conventional error-magnitude measures when the nonlinear interaction form is unknown. The implementation code and benchmark data are openly available in a persistent Zenodo repository.
\end{abstract}


\noindent\textbf{Software and benchmark package:} Zenodo -- \href{https://doi.org/10.5281/zenodo.23330544}{10.5281/zenodo.23330544}.

\section{Introduction}
Many scientific and engineering problems involve two or more simultaneously observed time series for which simple association is not the quantity of primary interest. A stronger and more operational question is whether the past of a candidate series contains information that improves prediction of another series beyond what is already contained in the target's own history. This distinction separates contemporaneous association from \emph{incremental predictive contribution}. It is central to directional time-series analysis in economics and finance, climate and hydrology, epidemiology, neuroscience and physiology, ecology, and complex dynamical systems, where strongly correlated signals may arise from synchronization, common forcing, seasonality, or shared trends without providing independent predictive information \cite{shojaie2021,edinburgh2021,sugihara2012,maher2016,bonotto2022,isufaj2025}.

Let $x_t$ denote a target time series and $y_t$ a candidate auxiliary series. The question studied here is
\begin{quote}
Does adding the aligned history of $y$ improve, or systematically alter, the out-of-sample prediction of future $x$ relative to using the history of $x$ alone?
\end{quote}
This formulation naturally leads to a matched restricted-versus-augmented prediction experiment: the two branches forecast exactly the same future target values, use the same training/test protocol and model class, and differ only in whether the history of $y$ is available. Pairwise Pearson or Spearman correlation remains useful descriptively, but it does not condition on the target history and does not directly answer this incremental forecasting question.

The canonical predictive framework is Granger causality. In its classical linear form, a restricted autoregression based on the past of $x$ is compared with a full model that additionally includes the past of $y$; a reduction in residual uncertainty is interpreted as evidence that $y$ is Granger-predictive for $x$ \cite{shojaie2021,stokes2017}. The same directional logic has been generalized far beyond linear autoregression. Kernel and nonlinear extensions, polynomial and Volterra expansions, neural networks, and other machine-learning predictors replace the fixed linear basis while retaining the restricted-versus-full comparison \cite{tank2021,rosol2022,hasanzadeh2021,wismuller2021}. These methods can detect nonlinear dependence that linear Granger models miss, but their response is inevitably tied to the selected predictor representation, lag structure, regularization, and model complexity. A correctly specified low-dimensional basis can be extremely sensitive, whereas a misspecified basis can fail even when a strong predictive channel is present.

A closely related but more model-agnostic strategy is to evaluate the forecasts directly on held-out data. Baseline and augmented models can be compared using changes in RMSE, MAE, MSE, forecast ratios, or out-of-sample $R^2$. This produces an interpretable effect size---how much forecast accuracy is gained by adding the candidate source---without requiring the statistic itself to assume a particular linear or nonlinear dependence. Statistical forecast-comparison procedures, most prominently the Diebold--Mariano test and its extensions, address a complementary question: whether the observed loss difference is statistically distinguishable from zero rather than how the predictive contribution should be represented \cite{diebold1995,grant2026}. Importantly, multi-horizon forecast evaluation is itself well established; the novelty of the present work is therefore not the use of several forecast horizons alone.

A second major family takes a state-space rather than restricted-versus-full forecasting view. Convergent cross mapping (CCM) asks whether information about a putative driver can be reconstructed from the delay-coordinate manifold of the affected variable and whether cross-map skill improves as the library size increases \cite{sugihara2012,chang2017}. Empirical dynamic modeling (EDM), simplex projection, and S-map methods extend this state-space viewpoint to nonlinear forecasting and state-dependent interaction strength \cite{deyle2016,chang2017}. These tools have been influential in ecology, epidemiology, climate and hydrology, physiology, neuroscience, and, more recently, financial time series \cite{sugihara2012,maher2016,bonotto2022,isufaj2025}. Their physical question is nevertheless different from incremental forecast gain: reconstructability of a source from an observed manifold need not be equivalent to improvement of a target forecast after the source is supplied. Standard CCM is also known to require careful treatment of embedding, convergence, noise, synchronization, common forcing, and short or discontinuous records \cite{monster2017,edinburgh2021}.

Table~\ref{tab:intro_methods} summarizes the principal method families relevant to the present study. The comparison is intentionally conceptual rather than hierarchical because the methods estimate different objects. Granger methods quantify restricted-versus-full predictive dependence; forecast-gain measures quantify held-out accuracy improvement; Diebold--Mariano-type procedures quantify evidence for a loss difference; CCM/EDM quantify state-space reconstruction or nonlinear state-dependent prediction. The proposed FEDE construction belongs to the predictive-comparison side of this landscape but uses a different response variable: the \emph{horizon evolution of the complete signed forecast-error distribution}.

\begin{table}[H]
\centering
\scriptsize
\renewcommand{\arraystretch}{1.15}
\caption{Representative method families for directional or incremental dependence between time series. The table emphasizes the object each method measures rather than treating the methods as interchangeable causal estimators.}
\label{tab:intro_methods}
\begin{tabularx}{\textwidth}{@{}p{2.35cm}X p{2.45cm} p{2.10cm} X p{2.15cm}@{}}
\toprule
Method family & Core question & Typical output & Role of forecast horizon & Main limitation for the present problem & Representative references \\
\midrule
Linear Granger causality & Does past $y$ reduce the prediction-error variance of $x$ beyond past $x$ in a linear model? & $F$/Wald statistic, $p$ value, GC index, variance ratio & Usually one-step or lag-model based & Misses nonlinear interactions when the linear basis is inadequate; sensitive to lag/model specification & \cite{shojaie2021,stokes2017} \\
\addlinespace
Nonlinear / ML Granger & Same restricted-versus-full question with a nonlinear predictor or feature basis & Nonlinear GC score, generalized variance, variable importance, test statistic & Usually one-step/few-step; model dependent & Sensitivity depends on architecture or chosen basis; greater tuning burden & \cite{tank2021,rosol2022,hasanzadeh2021,wismuller2021} \\
\addlinespace
Out-of-sample forecast gain & How much does adding $y$ improve held-out prediction of future $x$? & $\Delta$RMSE, $\Delta$MAE, loss ratio, OOS $R^2$ & Chosen explicitly by analyst; may be single- or multi-horizon & Collapses forecast behavior to an accuracy summary and does not describe how the error distribution changes with $h$ & \cite{diebold1995,grant2026} \\
\addlinespace
Diebold--Mariano-type forecast comparison & Is the mean loss difference between two forecast systems distinguishable from zero? & DM statistic and $p$ value & Horizon-specific or extended to multiple horizons & Primarily a significance test, not a coupling-magnitude or error-structure descriptor & \cite{diebold1995,grant2026} \\
\addlinespace
CCM / EDM & Is information about one variable recoverable from the reconstructed state of another, and does reconstruction converge? & Cross-map skill $\rho(L)$, convergence, nonlinear forecast skill, local coefficients & Library size and embedding are central; forecast horizon enters EDM variants & Reconstruction and predictive contribution are not equivalent; noise, synchronization, embedding, and common forcing matter & \cite{sugihara2012,chang2017,deyle2016,monster2017} \\
\addlinespace
\textbf{FEDE predictive-causality framework (this work)} & How does adding $y$ change the \emph{distributional evolution} of out-of-sample forecast errors for $x$ across horizons? & Directed FEDE $\Delta$ response and early-horizon entropy-slope response & \textbf{Explicitly uses $h=1,\ldots,H$ as the analysis axis} & Model-relative; requires multi-horizon forecasts and an adequate predictive engine & Present study \\
\bottomrule
\end{tabularx}
\end{table}

Forecasting in nonlinear dynamical systems provides a natural setting in which the horizon dependence itself carries information. Classical nonlinear prediction and state-space reconstruction showed that the degradation of forecast skill with lead time reflects underlying dynamical structure \cite{farmer1987,casdagli1989,sugihara1990,takens1981}. Our earlier work used machine-learning forecast-error growth to estimate largest-Lyapunov-exponent-related behavior from scalar chaotic time series \cite{velichko2025chaos}. That line of work suggests that a forecast should not always be reduced to a single error number: the \emph{shape} of the error evolution over $h$ can be an informative observable.

The gap addressed here is therefore narrower, but also more specific, than the general problem of detecting directionality. We do not propose a new definition of structural or physical causality and do not claim novelty for the restricted-versus-augmented forecast comparison. Instead, FEDE treats the distribution of signed out-of-sample forecast errors at each horizon as the primary object. For a target-only forecast and a matched target-plus-auxiliary forecast, Shannon entropy is computed from the signed error distributions over $h=1,\ldots,H$; the change of the resulting entropy profile between short and longer horizons becomes a scalar, directed, model-relative descriptor. We refer to the complete construction as the \emph{FEDE predictive-causality framework}. Throughout this work, predictive causality is used strictly in the Granger operational sense: a candidate source is predictive-causal for a target only insofar as its past provides additional out-of-sample predictive information beyond the target's own past, within the specified model and conditioning set. The question is therefore no longer only whether $y$ decreases an average loss, but whether access to $y$ changes the way predictive uncertainty is redistributed across the forecast horizon.

This distinction motivates the controlled benchmarks used in the present study. A hidden nonlinear source is first constructed so that the raw auxiliary series can have essentially zero linear association with the future target even though a genuine predictive channel is present. The analysis then tests whether a flexible predictor can extract this hidden information and whether FEDE gives a more reproducible coupling response than conventional mean forecast gain. Direction and source controls ask whether the response is specific to the true $Y\rightarrow X$ channel rather than to the reverse direction or to unrelated structured auxiliaries. A second benchmark varies the hidden functional form while keeping its power normalized, allowing direct comparison with chronological out-of-sample linear and polynomial Granger-style scores and with basic CCM convergence. A final predefined parameter grid then varies the FEDE analysis scales $W$, $m$, and $H$ at fixed noise to test whether the nonlinear dose response survives changes in the method's own window, history, and horizon settings.

The main contributions of the study are fivefold. First, we formulate the FEDE predictive-causality framework as a matched multi-horizon construction in which the only information difference between two forecast branches is the availability of the candidate auxiliary history. Second, we introduce FEDE-based directed descriptors that quantify changes in the entropy structure of signed forecast errors across horizons rather than only changes in mean error magnitude. Third, we validate source/direction specificity and robustness to several hidden nonlinear functional forms, including cases in which raw linear association is deliberately suppressed. Fourth, we compare the FEDE response with conventional multi-horizon forecast gain, chronological out-of-sample linear and degree-2 polynomial Granger-style controls, and CCM, while separately reporting computational cost. Fifth, we test the stability of the FEDE dose response over a predefined $3\times3\times3$ grid of window lengths, history lengths, and maximum forecast horizons. The comparison is designed to identify scope rather than to assert universal superiority: a correctly specified parametric model is expected to outperform a flexible generic method when the true interaction basis is known a priori.

The remainder of the paper is organized as follows. Section~2 defines the matched forecasting construction, FEG/FEDE descriptors, dose--response evaluation, directed contribution, time-resolved evaluation, and the competing Granger/CCM estimators. Section~3 separates the controlled tests into a linear-reference benchmark family (L1--L2) and a nonlinear Gaussian-AR(1) benchmark family (N1--N5), and states explicitly how each benchmark maps to the corresponding results. Section~4 presents calibration, detection, nonlinear extraction, specificity, method-comparison, tracking, computational, and parameter-robustness results. Section~5 discusses interpretation, model dependence, comparison with established directional methods, computational cost, and limitations. Section~6 summarizes the conclusions.

\section{Methodology}
\subsection{Matched predictive-contribution construction}
For history length $m$, define
\begin{align}
\mathbf{x}_t &= [x_{t-m+1},\ldots,x_t],\\
\mathbf{z}_t &= [x_{t-m+1},\ldots,x_t,\,y_{t-m+1},\ldots,y_t].
\end{align}
Two matched predictors are fit using the same model class, hyperparameters, training indices, test indices, and forecast horizons $h=1,\ldots,H$:
\begin{align}
\widehat{x}^{(x)}_{t+h} &= F_h(\mathbf{x}_t),\\
\widehat{x}^{(x,y)}_{t+h} &= F_h(\mathbf{z}_t).
\end{align}
The corresponding signed errors are
\begin{align}
e_x(i,h)&=x_{i+h}-\widehat{x}^{(x)}_{i+h},\\
e_{xy}(i,h)&=x_{i+h}-\widehat{x}^{(x,y)}_{i+h}.
\end{align}
Any systematic difference between these two matched error sets is therefore attributable to the availability of the auxiliary history within the chosen prediction model.

\begin{figure}[H]
\centering
\resizebox{\textwidth}{!}{%
\begin{tikzpicture}[
  node distance=10mm and 13mm,
  every node/.style={font=\small},
  input/.style={draw,rounded corners,align=center,minimum width=31mm,minimum height=11mm,fill=blue!4},
  model/.style={draw,rounded corners,align=center,minimum width=34mm,minimum height=12mm,fill=gray!8},
  pred/.style={draw,rounded corners,align=center,minimum width=29mm,minimum height=11mm,fill=green!5},
  err/.style={draw,rounded corners,align=center,minimum width=28mm,minimum height=11mm,fill=orange!7},
  metric/.style={draw,rounded corners,align=center,minimum width=43mm,minimum height=19mm,fill=yellow!10},
  arr/.style={-{Latex[length=2.2mm]},thick}
]

\node[input] (xhist) {$x$ history\\$\mathbf{x}_t$};
\node[input, below=13mm of xhist] (xyhist) {$x+y$ history\\$\mathbf{z}_t$};

\node[model, right=16mm of xhist] (mb) {predictor $F$\\target-only};
\node[model, right=16mm of xyhist] (ma) {same predictor $F$\\target+auxiliary};

\node[pred, right=15mm of mb] (pb) {$\widehat{x}^{(x)}_{t+h}$\\$h=1,\ldots,H$};
\node[pred, right=15mm of ma] (pa) {$\widehat{x}^{(x,y)}_{t+h}$\\$h=1,\ldots,H$};

\node[err, right=15mm of pb] (eb) {$e_x(i,h)$};
\node[err, right=15mm of pa] (ea) {$e_{xy}(i,h)$};

\node[metric, right=16mm of $(eb)!0.5!(ea)$] (metrics) {matched comparison\\\vspace{1mm}$G_{\log\RMSE}$, $G_{\rm profile}$, $\Delta\lambda$,\\$D_{\FEDE}$, $E_{\FEDE}$};

\draw[arr] (xhist) -- (mb);
\draw[arr] (xyhist) -- (ma);
\draw[arr] (mb) -- (pb);
\draw[arr] (ma) -- (pa);
\draw[arr] (pb) -- (eb);
\draw[arr] (pa) -- (ea);
\draw[arr] (eb) -- (metrics);
\draw[arr] (ea) -- (metrics);

\node[draw,dashed,rounded corners,fit=(mb)(ma),inner sep=4mm,label={[font=\scriptsize]above:same model class, hyperparameters, train/test split}] {};
\end{tikzpicture}}
\caption{Core construction for auxiliary predictive contribution. The two branches forecast exactly the same future target values. The only information difference is the availability of the aligned auxiliary history $y$. Conventional accuracy gain and FEG/FEDE descriptors are all computed from this same matched pair of out-of-sample forecasts.}
\label{fig:framework}
\end{figure}
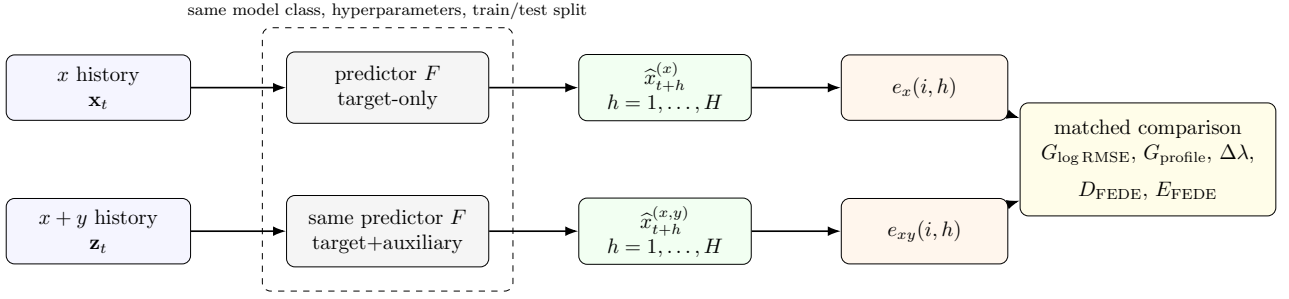

Within the FEDE predictive-causality framework, this comparison is model-relative: predictive causality refers only to additional out-of-sample predictive information supplied by the candidate source. It should not be interpreted as structural or physical causal influence without additional assumptions, conditioning, or controls.

\subsection{Conventional predictive gain}
For each horizon,
\begin{align}
\RMSE_x(h)&=\sqrt{\frac{1}{N_h}\sum_i e_x(i,h)^2},\\
\RMSE_{xy}(h)&=\sqrt{\frac{1}{N_h}\sum_i e_{xy}(i,h)^2}.
\end{align}
The main conventional summary is the mean multi-horizon log-RMSE gain,
\begin{equation}
G_{\log\RMSE}=\frac{1}{H}\sum_{h=1}^{H}
\log\frac{\RMSE_x(h)+\varepsilon}{\RMSE_{xy}(h)+\varepsilon}.
\label{eq:logrmse}
\end{equation}
Positive values indicate lower forecast error when the auxiliary history is included.

\subsection{Forecast-error growth (FEG) profile}
For either matched error matrix $e(i,h)$, define
\begin{equation}
g(h)=\frac{1}{N_h}\sum_{i=1}^{N_h}\log\left(\max(|e(i,h)|,\varepsilon)\right).
\label{eq:feg}
\end{equation}
The target-only and target-plus-auxiliary profiles are denoted $g_x(h)$ and $g_{xy}(h)$. Their mean separation is
\begin{equation}
G_{\mathrm{profile}}=\frac{1}{H}\sum_{h=1}^{H}\left[g_x(h)-g_{xy}(h)\right].
\end{equation}
A linear fit $g(h)\approx c+\lambda h$ also gives the horizon-growth response
\begin{equation}
\Delta\lambda=\lambda_x-\lambda_{xy}.
\end{equation}

\subsection{Forecast-error distribution entropy (FEDE)}
FEDE describes the distribution of signed errors at each forecast horizon. For every matched window, baseline and auxiliary-informed errors are pooled to define one common histogram range $[e_{\min},e_{\max}]$. The same $B=16$ bins are then used for both branches and all horizons. If $p_j(h)$ denotes the empirical probability in bin $j$,
\begin{equation}
\FEDE(h)=-\frac{\sum_{j=1}^{B}p_j(h)\log p_j(h)}{\log B},
\qquad 0\leq\FEDE(h)\leq1.
\label{eq:fede}
\end{equation}
We use two horizon-structure summaries:
\begin{align}
\Delta\FEDE &= \FEDE(H)-\FEDE(1),\\
s^{\mathrm{early}}_{\FEDE} &= \operatorname{slope}\!\left(\FEDE(h),h=1,\ldots,5\right).
\end{align}
Their matched responses are
\begin{align}
D_{\FEDE}&=\Delta\FEDE_{xy}-\Delta\FEDE_x,\\
E_{\FEDE}&=s^{\mathrm{early}}_{\FEDE,xy}-s^{\mathrm{early}}_{\FEDE,x}.
\end{align}
The second quantity, $E_{\FEDE}$, is the \emph{FEDE early-slope change}. It is particularly useful in the present linear benchmark because it combines weak-coupling sensitivity with an approximately linear calibration to the known coupling strength.

\subsection{Prediction engines}
Ridge regression is used as the linear predictive reference with $\alpha=10^{-6}$ \cite{hoerl1970}. LightGBM provides a nonlinear comparison with 50 estimators, 7 leaves, maximum depth 3, learning rate 0.05, minimum child size 5, and $\ell_2$ regularization 0.1 \cite{ke2017}. LightGBM uses one internal thread. The main figures use history $m=1$; $m=5$ is retained in the software as a sensitivity setting. The forecast horizon is $H=14$, and the chronological train/test split is 60:40.

\subsection{Dose--response evaluation: calibration, detection, and reproducibility}
The controlled benchmarks impose a known coupling strength $\gamma$, which provides an external calibration axis for any scalar response statistic $S$. In this paper, $S$ may denote a conventional forecast-gain quantity or an FEG/FEDE-derived response such as $G_{\log\RMSE}$, $G_{\mathrm{profile}}$, $\Delta\lambda$, $D_{\FEDE}$, or $E_{\FEDE}$. For a fixed noise level $\sigma$, let $\overline{S}(\gamma,\sigma)$ denote the mean of that statistic across the independent realizations at a given coupling level. Calibration linearity is assessed from
\begin{equation}
\overline{S}(\gamma,\sigma)\approx \alpha+\beta\gamma.
\label{eq:condition_calibration}
\end{equation}
The coefficient of determination $R^2_{S,\gamma}$ measures how closely the condition-mean response follows a linear scale in the known imposed coupling. This complements monotonicity: a statistic can preserve rank ordering while having a strongly nonlinear response to $\gamma$.

The exact null condition is $\gamma=0$. For each noise level and statistic, the $\gamma=0$ realizations define a one-sided 95th-percentile null threshold after orienting the statistic according to its dose--response direction. The detection rate at $\gamma>0$ is the fraction of realizations exceeding that threshold. For a required sustained detection probability $p$, the detection limit is
\begin{equation}
\gmin^{(p)}(\sigma;S)=
\min\left\{\gamma_j: p_{\mathrm{det}}(\gamma_k,\sigma;S)\ge p\ \forall k\ge j\right\}.
\label{eq:gmin}
\end{equation}
We report both $p=0.8$ and $p=0.9$ where the tested coupling range permits it. If no tested value satisfies Eq.~\eqref{eq:gmin}, the result is reported as exceeding the largest tested $\gamma$.

Condition-mean calibration alone does not penalize a descriptor whose response varies strongly between independent realizations. We therefore also fit the dose--response relation separately in every realization $r=1,\ldots,R$,
\begin{equation}
M_r(\gamma)=\alpha_r+\beta_r\gamma,
\label{eq:realization_slope}
\end{equation}
where $M_r$ is the realization-specific scalar predictive response. Let $\bar\beta$ denote the mean slope across the $R$ independent realizations and $s_\beta$ the sample standard deviation of these realization-specific slopes (estimated with the usual $R-1$ denominator). The standard error (SE) of the mean slope is
\begin{equation}
\mathrm{SE}(\bar\beta)=\frac{s_\beta}{\sqrt{R}},
\end{equation}
and the signed response-sensitivity statistic is
\begin{equation}
T=\frac{\bar\beta}{\mathrm{SE}(\bar\beta)}
=\frac{\bar\beta}{s_\beta/\sqrt{R}}.
\label{eq:T}
\end{equation}
Its magnitude $|T|$ quantifies the strength and reproducibility of the realization-wise dose--response slope relative to its between-realization uncertainty, while $\operatorname{sign}(T)$ gives the orientation of that response. Thus, a large negative $T$ is not intrinsically a low-sensitivity result: it can represent a strong and reproducible \emph{inverse} response. In the present metrics, however, the orientation is defined a priori so that positive log-RMSE gain and positive FEDE changes correspond to the expected direction of useful auxiliary predictive contribution. Hence a large positive $T$ is the desired response, $T\approx0$ indicates little reproducible dose response, and $T<0$ indicates an anti-aligned response. For metrics whose physically meaningful direction is not defined in advance, both $T$ and $|T|$ should be reported, or the metric orientation should be specified a priori before calculating $T$; the sign should not be discarded post hoc. Because both numerator and denominator scale identically, $T$ is invariant to a constant positive rescaling of the metric and can therefore compare log-RMSE gain with FEDE descriptors having different numerical units. The 95\% confidence interval for the mean slope is
\begin{equation}
\bar\beta\pm t_{0.975,R-1}\,\mathrm{SE}(\bar\beta).
\end{equation}
Equation~\eqref{eq:T} is algebraically the one-sample $t$ statistic for the realization-wise slopes, but here it is used descriptively as a noise-normalized sensitivity and reproducibility index rather than as a stand-alone multiple-testing significance claim.

\subsection{Directed predictive contribution and source specificity}
The matched construction is naturally directional because direction is defined by the forecast target rather than by the sign of the response. For a generic target series $u_t$ and a candidate source $v_t$, let $\Delta\FEDE_{u\rightarrow u}$ denote the FEDE delta obtained from a target-only forecast and let $\Delta\FEDE_{(u,v)\rightarrow u}$ denote the corresponding quantity after adding the candidate source. We define the directed FEDE contribution
\begin{equation}
D_{\FEDE}^{v\rightarrow u}
=
\Delta\FEDE_{(u,v)\rightarrow u}
-
\Delta\FEDE_{u\rightarrow u}.
\label{eq:directed_fede}
\end{equation}
Thus, testing $Y\rightarrow X$ compares $X\rightarrow X$ with $(X,Y)\rightarrow X$, whereas testing the reverse direction $X\rightarrow Y$ compares $Y\rightarrow Y$ with $(Y,X)\rightarrow Y$. A positive or negative value of $D_{\FEDE}^{v\rightarrow u}$ describes the orientation of the predictive response for the chosen direction; the sign itself does not define the direction.

For controlled specificity tests, the same candidate-source comparison can be repeated with unrelated auxiliary signals while keeping the target, prediction engine, windowing, and coupling scan unchanged. If the response is source-specific, the true driver should show a reproducible dependence on the imposed coupling strength, whereas unrelated candidate sources should remain near their null response. For each candidate direction/source, realization-wise slopes of $D_{\FEDE}^{v\rightarrow u}(\gamma)$ are summarized by the signed statistic $T$ in Eq.~\eqref{eq:T}. The resulting quantity is a test of \emph{direction-specific incremental predictability} relative to a specified model class. It is not, by itself, a claim of structural or interventional causality; common drivers, feedback, omitted state variables, or model misspecification can in general generate predictive asymmetries without establishing a causal mechanism.

\subsection{Competing directional estimators}
The comparative benchmark uses each competing method in a form natural to its own construction rather than forcing every method onto the FEDE horizon axis. This distinction is important: FEDE and mean log-RMSE use the full forecast range $h=1,\ldots,H$, the Granger-style controls are one-step predictive comparisons, and CCM is summarized by convergence with library size.

\paragraph{Chronological out-of-sample Granger-style gain.}
Classical Granger analysis compares a restricted model of the target from its own past with a full model that also contains the candidate source past \cite{shojaie2021}. To obtain an effect-size comparator without giving the larger full model an automatic in-sample residual advantage, we use chronological held-out scoring rather than the conventional in-sample $F$ test. In every $W=100$ window and for history $p=1$, the supervised rows are split chronologically into 60\% training and 40\% test data. The restricted and full linear models are
\begin{align}
x_{t+1} &= a_0+a_1x_t+\varepsilon^{(R)}_{t+1},\\
x_{t+1} &= c_0+c_1x_t+d_1y_t+\varepsilon^{(F)}_{t+1}.
\end{align}
Standardization parameters are fitted on the training rows only. Both models are estimated by ordinary least squares on the training subset and evaluated on the same held-out target values. The reported directional effect size is
\begin{equation}
G_{\mathrm{GC,OOS}}^{y\rightarrow x}
=
\log\frac{\mathrm{MSE}^{\mathrm{test}}_{R}}
          {\mathrm{MSE}^{\mathrm{test}}_{F}}.
\label{eq:gc_oos}
\end{equation}
Thus $G_{\mathrm{GC,OOS}}>0$ means that adding $y_t$ improves unseen one-step prediction, whereas a negative value means that the additional source degrades generalization.

The nonlinear Granger-style comparator retains exactly the same train/test protocol but applies a fixed degree-2 polynomial expansion. For $p=1$, the restricted model receives $x_t$ and $x_t^2$, while the full model receives
\begin{equation}
x_t,\;y_t,\;x_t^2,\;x_ty_t,\;y_t^2.
\end{equation}
The degree is deliberately fixed at two for \emph{all} hidden-channel forms; it is not retuned after the true nonlinearity is changed. This makes the quadratic channel a matched-basis case and the remaining channels tests of representation mismatch, consistent with the broader nonlinear Granger literature in which polynomial/Volterra and machine-learning representations define the class of interactions that can be extracted \cite{hasanzadeh2021,rosol2022,tank2021}.

\paragraph{Convergent cross mapping (CCM).}
For a candidate direction $y\rightarrow x$, convergent cross mapping (CCM) does not forecast future $x$ from past $y$. Instead, it asks whether the contemporaneous source value $y_t$ can be reconstructed from nearby states on the delay-coordinate manifold built from $x$ \cite{sugihara2012}. The apparent reversal is intrinsic to cross mapping: if $Y$ contributes to the dynamics of $X$, information about $Y$ may be encoded in the reconstructed state of $X$. In each $W=100$ window, the delay is fixed at $\tau=1$. The embedding dimension $E\in\{1,2,3,4\}$ is selected by the one-step nearest-neighbor prediction skill of the target, and source reconstruction uses the $E+1$ nearest neighbors in the target manifold. Cross-map skill is the Pearson correlation $\rho(L)$ between the observed source values and their nearest-neighbor reconstructions from the target manifold using library size $L$. Three random library draws are averaged at
\begin{equation}
L\in\{30,50,70,90\},
\end{equation}
and the scalar convergence comparator is
\begin{equation}
M_{\mathrm{CCM}}=\Delta\rho=\rho(90)-\rho(30).
\label{eq:ccm_delta}
\end{equation}
Thus CCM is a state-space cross-reconstruction test rather than a future-forecast comparison. The reverse direction exchanges source and target. This is a deliberately basic convergence implementation: it does not include lag scanning, surrogate testing, or multivariate conditioning, and its interpretation is therefore restricted to the present controlled benchmark \cite{monster2017}.

\begin{table}[H]
\centering
\scriptsize
\caption{Scalar quantities used in the method-comparison benchmark. Each method is evaluated on the same $W=100$ windows and the same controlled $\gamma$ scan, but its native within-window construction is preserved.}
\label{tab:method_objects}
\begin{tabular}{llll}
\toprule
Method & Within-window object & Scalar response & Native axis \\
\midrule
FEDE $\Delta$ + LightGBM & signed-error entropy profile & $D_{\FEDE}$ & $h=1,\ldots,14$ \\
Mean log-RMSE + LightGBM & forecast-error magnitude & $G_{\log\RMSE}$ & $h=1,\ldots,14$ \\
Linear Granger-style OOS & restricted vs full OLS & Eq.~\eqref{eq:gc_oos} & one step \\
Polynomial Granger-style OOS & restricted vs degree-2 full OLS & Eq.~\eqref{eq:gc_oos} & one step \\
CCM & source reconstruction from target manifold & $\rho(90)-\rho(30)$ & library size \\
\bottomrule
\end{tabular}
\end{table}

\subsection{Time-resolved evaluation}
For the intermittent nonlinear benchmark N2, the same matched forecasts are evaluated in rolling windows. For a window ending at time $\tau$, $\bar\gamma(\tau)$ is the mean imposed coupling inside that window. Because a finite rolling window overlaps a switching boundary, $\bar\gamma(\tau)$ provides the appropriate window-level ground truth rather than the instantaneous value at the window end.

For presentation and comparison with the synthetic nonlinear oracle defined with the nonlinear benchmark family in Section~\ref{sec:nonlinear_common}, rolling metrics are averaged across the independent realizations at each window end. If $M_r(\tau)$ is a metric and $O_r(\tau)$ is the oracle in realization $r$, then
\begin{align}
\overline{M}(\tau)&=R^{-1}\sum_r M_r(\tau),\\
\overline{O}(\tau)&=R^{-1}\sum_r O_r(\tau).
\end{align}
Time-resolved tracking quality is quantified by both Spearman $\rho_S$ and Pearson $r_P$. We report correlations of the \emph{unsmoothed} ensemble-mean response $\overline{M}(\tau)$ with (i) the ensemble-mean nonlinear oracle $\overline{O}(\tau)$ and (ii) the known window-mean imposed coupling $\bar\gamma(\tau)$. The second comparison is the most direct benchmark of temporal tracking because $\bar\gamma(\tau)$ is known exactly by construction. Negative predictive responses are retained numerically in all calculations; in the time-domain FEDE figures they are shown with lower opacity rather than clipped. The displayed N2 curves are unsmoothed ensemble means without confidence bands so that the window-induced temporal broadening remains visible directly.

\section{Controlled synthetic experiments}
The synthetic study is organized into two deliberately distinct benchmark families. The \emph{linear-reference} family L1--L2 uses separately initialized, decorrelated chaotic logistic-map series and asks whether the proposed descriptors recover a known direct linear predictive contribution. The \emph{nonlinear} family N1--N5 uses Gaussian AR(1) series and deliberately hides the tested source contribution behind nonlinear transformations that are not supplied to the predictors. The symbols $y$ and $q$ denote the tested source and an unobserved background forcing, respectively, in both families, but their generating processes are deliberately different. Table~\ref{tab:benchmark_map} provides the roadmap from each benchmark to its scientific purpose and corresponding results.

\begin{table}[H]
\centering
\small
\caption{Roadmap of the controlled synthetic benchmarks. L1--L2 and N1--N5 are separate generating families; they share the same matched forecasting logic but not the same source/background process.}
\label{tab:benchmark_map}
\begin{tabularx}{\textwidth}{@{}l>{\raggedright\arraybackslash}p{31mm}>{\raggedright\arraybackslash}X>{\raggedright\arraybackslash}p{28mm}@{}}
\toprule
Benchmark & Generating family & Main purpose & Results \\
\midrule
L1 & logistic-map $y,q$ & calibration to known direct linear coupling & Sec.~\ref{sec:res_expA_calibration} \\
L2 & logistic-map $y,q$ & weak coupling, process noise, and null-relative detection & Secs.~\ref{sec:res_expB_weak}--\ref{sec:res_expB_detection} \\
N1 & Gaussian AR(1) $y,q$ & hidden nonlinear extraction and Ridge--LightGBM comparison & Secs.~\ref{sec:res_n1_validation}--\ref{sec:res_n1_extract} \\
N2 & Gaussian AR(1) $y,q$ & time-resolved tracking of intermittent coupling & Sec.~\ref{sec:res_n2_intermittent} \\
N3 & Gaussian AR(1) $y,q$ & direction and source specificity & Sec.~\ref{sec:res_n3_specificity} \\
N4 & Gaussian AR(1) $y,q$ & robustness to hidden functional form; Granger/CCM comparison & Secs.~\ref{sec:res_n4_methods}, \ref{sec:res_n4_timing} \\
N5 & Gaussian AR(1) $y,q$ & robustness to $W$, $m$, and $H$ & Sec.~\ref{sec:res_n5_robustness} \\
\bottomrule
\end{tabularx}
\end{table}

\subsection{Linear-reference benchmark family: common generating system}\label{sec:linear_family}
Benchmarks L1 and L2 use two separately initialized deterministic chaotic drivers, $y_n$ and $q_n$, generated from the fully chaotic logistic map
\begin{equation}
z_{n+1}=4z_n(1-z_n),
\label{eq:linear_logistic}
\end{equation}
where $z$ denotes either driver and the two sequences use separate initial conditions. Both are standardized after a transient. Candidate realizations are accepted only when the absolute retained correlation between the two drivers is below 0.15. The target is
\begin{equation}
x_{n+1}=0.55x_n+0.20q_n+0.60\gamma y_n+\sigma\epsilon_n,
\qquad \epsilon_n\sim\mathcal{N}(0,1).
\label{eq:system}
\end{equation}
Here $y_n$ is the tested auxiliary source available to the augmented predictor, whereas $q_n$ is an independent unobserved background chaotic forcing that contributes target variability unavailable to either forecasting branch. The parameter $\gamma$ is the known strength of the added direct linear predictive channel $y_n\rightarrow x_{n+1}$. This family therefore provides a reference case in which the injected source contribution is linear and Ridge is a natural baseline. It is used to ask three elementary questions before introducing hidden nonlinear structure: whether a response is calibrated to known $\gamma$, how early weak coupling separates from the exact null $\gamma=0$, and how process noise changes those properties.

\subsection{Benchmark L1: stationary linear calibration}\label{sec:expA_method}
Benchmark L1 uses $\sigma=0$, 5000 retained samples per realization, 10 independent realizations, $W=100$, step 100, and
\begin{equation}
\gamma\in\{0,0.01,0.02,0.03,0.05,0.075,0.10,0.15,0.20,0.30,0.50,0.70,1.00\}.
\end{equation}
Each run has a constant $\gamma$. The benchmark therefore asks how each predictive metric changes as the known linear coupling strength is increased across otherwise comparable runs. Calibration linearity is evaluated with Eq.~\eqref{eq:condition_calibration}. The corresponding results are reported in Section~\ref{sec:res_expA_calibration}.

\subsection{Benchmark L2: weak linear coupling under process noise}\label{sec:expB_method}
Benchmark L2 uses 30 independent realizations, 3000 retained samples per realization, $W=100$, step 100,
\begin{equation}
\gamma\in\{0,0.01,0.02,0.03,0.05,0.075,0.10,0.15,0.20,0.30\},
\end{equation}
and
\begin{equation}
\sigma\in\{0,0.05,0.10,0.20,0.40,0.80\}.
\end{equation}
For every fixed realization and noise level, the same logistic-map $y$, logistic-map $q$, and Gaussian process-noise vector are reused across all $\gamma$ values. This paired design reduces unrelated variation in the dose--response comparison. Benchmark L2 is used for the weak-coupling zoom, the noise-dependent calibration study, and the null-relative detection limit in Eq.~\eqref{eq:gmin}. Results are reported in Sections~\ref{sec:res_expB_weak}--\ref{sec:res_expB_detection}.

\begin{table}[H]
\centering
\caption{Main settings of the linear-reference benchmarks L1 and L2.}
\label{tab:protocol}
\begin{tabular}{lcc}
\toprule
Setting & Benchmark L1 & Benchmark L2 \\
\midrule
Purpose & calibration & weak coupling + noise \\
Driver family & logistic map & logistic map \\
Realizations & 10 & 30 \\
Retained samples & 5000 & 3000 \\
Window length $W$ & 100 & 100 \\
Forecast horizon $H$ & 14 & 14 \\
Main history $m$ & 1 & 1 \\
Coupling $\gamma$ & 0--1.0 & 0--0.30 \\
Noise $\sigma$ & 0 & 0--0.80 \\
Predictors & Ridge, LightGBM & Ridge, LightGBM \\
\bottomrule
\end{tabular}
\end{table}

\subsection{Nonlinear benchmark family: common Gaussian-AR(1) generating system}\label{sec:nonlinear_common}
Benchmarks N1--N5 use a different generating family from L1--L2. The tested auxiliary source $y_t$ is a stationary zero-mean, unit-variance Gaussian AR(1) process,
\begin{equation}
y_{t+1}=\varphi_y y_t+\sqrt{1-\varphi_y^2}\,\eta_t^{(y)},
\qquad \varphi_y=0.60,
\label{eq:ar1_y}
\end{equation}
and the target is also driven by an independent unobserved background Gaussian AR(1) forcing,
\begin{equation}
q_{t+1}=\varphi_q q_t+\sqrt{1-\varphi_q^2}\,\eta_t^{(q)},
\qquad \varphi_q=0.35,
\label{eq:ar1_q}
\end{equation}
with independent standard-normal innovations. The innovation sequences $\eta_t^{(y)}$ and $\eta_t^{(q)}$ generate the stochastic AR(1) source and background processes themselves; they are distinct from the independent target process noise $\epsilon_t$ in Eq.~\eqref{eq:nonlinear_system}, whose contribution is explicitly controlled by $\sigma$. Realizations are regenerated until the retained absolute correlation $|r(y,q)|$ is at most 0.15.

The term \emph{nonlinear benchmark} refers to the hidden $Y\rightarrow X$ interaction, not to the marginal source processes: $y_t$ and $q_t$ are linear Gaussian AR(1) processes, whereas the tested predictive channel enters through a nonlinear transformation of $y_t$. The tested contribution is intentionally hidden from ordinary raw linear association. Rather than entering the target through $y_t$ itself, the source acts through the normalized even transformation
\begin{equation}
\phi_2(y)=\frac{y^2-1}{\sqrt{2}}.
\label{eq:phi}
\end{equation}
For a standard Gaussian variable, $\phi_2(y)$ has zero mean and unit variance and is orthogonal to the raw first-order component $y$ in the population, so $\operatorname{Corr}[y,\phi_2(y)]=0$ theoretically. To preserve this property in finite records, candidate realizations are additionally required to satisfy $|r(y,\phi_2(y))|\le0.05$. The common stationary nonlinear target is
\begin{equation}
x_{t+1}=0.55x_t+0.20q_t+0.60\gamma\,\phi_2(y_t)+\sigma\epsilon_t,
\qquad \epsilon_t\sim\mathcal{N}(0,1).
\label{eq:nonlinear_system}
\end{equation}
Thus the nonlinear benchmarks contain a genuine, controlled predictive channel $Y\rightarrow X$, but its imposed contribution is deliberately difficult to detect from the raw linear association between $y_t$ and future $x$. Crucially, neither predictor is given $\phi_2(y_t)$: Ridge and LightGBM receive only the raw history of $y_t$ in the augmented branch. Neither branch receives $q_t$, $\eta_t^{(y)}$, $\eta_t^{(q)}$, or $\epsilon_t$. The comparison therefore tests whether the chosen prediction engine can extract the hidden nonlinear information from raw $y$ before FEDE summarizes how that information changes the forecast-error distribution across horizons. The background process $q_t$ serves only as independent temporally structured forcing of the target.

Because the generating mechanism is known here, a diagnostic synthetic oracle is available:
\begin{equation}
O=r\!\left(\phi_2(y_t),x_{t+1}\right),
\label{eq:oracle}
\end{equation}
computed on the same window as the predictive metrics. The corresponding raw diagnostic is $r(y_t,x_{t+1})$. The oracle is not an estimator intended for real data and is not a competitor to FEDE; it is a validation device that checks whether the imposed nonlinear source contribution is present even when raw linear association is suppressed.

\subsection{Benchmark N1: stationary hidden nonlinear coupling}\label{sec:n1_method}
Benchmark N1 uses the common nonlinear system in Eqs.~\eqref{eq:ar1_y}--\eqref{eq:nonlinear_system} with 20 independent realizations, 3000 retained samples, $W=100$, stationary step 100, $H=14$, and histories $m\in\{1,5\}$. The coupling grid is
\begin{equation}
\gamma\in\{0,0.01,0.02,0.03,0.05,0.075,0.10,0.15,0.20,0.30,0.50,0.70,1.00\},
\end{equation}
and the noise grid is
\begin{equation}
\sigma\in\{0,0.05,0.10,0.20\}.
\end{equation}
For a fixed realization and noise level, the same AR(1) source $y$, background forcing $q$, and Gaussian noise realization are reused across all $\gamma$ values. Thus the dose--response comparison is paired with respect to the stochastic forcing. Benchmark N1 first verifies that the oracle grows with $\gamma$ while the raw correlation remains near zero, and then compares whether Ridge and LightGBM can extract the hidden channel from raw $y$. Results are reported in Sections~\ref{sec:res_n1_validation}--\ref{sec:res_n1_extract}.

\subsection{Benchmark N2: intermittent hidden nonlinear coupling}\label{sec:n2_method}
Benchmark N2 uses the same Gaussian-AR(1) generating system and hidden quadratic channel, but the true coupling follows the rectangular schedule
\begin{equation}
\gamma(t):\quad 0\rightarrow0.05\rightarrow0\rightarrow0.10\rightarrow0\rightarrow0.20\rightarrow0\rightarrow0.40.
\label{eq:schedule}
\end{equation}
Each of the eight blocks lasts 1000 samples, giving 8000 retained samples in total. Ten independent realizations are evaluated at $\sigma=0.10$, history $m=1$, maximum forecast horizon $H=14$, and rolling step 20. Two window lengths are analyzed on the same realizations: $W=200$ and $W=400$. The $W=400$ result is used for the main-text visualization because the longer window gives a more stable response estimate, while the $W=200$ result is retained in the Appendix as a higher-temporal-resolution comparison. The exact rectangular schedule is shown as the ground truth in the time-domain figures; rolling oracle and predictive quantities remain window-based and therefore broaden transitions over the finite temporal support of each window. Benchmark N2 therefore tests not only whether the hidden nonlinear contribution can be followed in time, but also the practical stability--temporal-resolution trade-off of the rolling analysis. Results are reported in Section~\ref{sec:res_n2_intermittent}.

\begin{table}[H]
\centering
\caption{Settings of the stationary and intermittent hidden nonlinear benchmarks N1 and N2.}
\label{tab:nonlinear_protocol}
\begin{tabular}{lcc}
\toprule
Setting & N1 stationary & N2 intermittent \\
\midrule
Tested source & Gaussian AR(1), $\varphi_y=0.60$ & same \\
Background forcing & Gaussian AR(1), $\varphi_q=0.35$ & same \\
Hidden transform & $\phi_2(y)=(y^2-1)/\sqrt{2}$ & same \\
Predictor input & raw $y$ only & raw $y$ only \\
Realizations & 20 & 10 \\
Retained samples & 3000 & 8000 \\
Window length $W$ & 100 & 200, 400 \\
Rolling step & 100 & 20 \\
Forecast horizon $H$ & 14 & 14 \\
Histories & 1, 5 & 1 \\
Noise $\sigma$ & 0, 0.05, 0.10, 0.20 & 0.10 \\
Coupling & 13 stationary levels, 0--1 & Eq.~\eqref{eq:schedule}; 1000 samples/level \\
\bottomrule
\end{tabular}
\end{table}

\subsection{Benchmark N3: directional and source specificity}\label{sec:n3_method}
Benchmark N3 retains the N1 stationary Gaussian-AR(1) generating system, the full 13-point $\gamma$ grid, $\sigma\in\{0,0.05,0.10,0.20\}$, 20 independent realizations, 3000 retained samples, $W=100$, step 100, $H=14$, and history $m=1$. Only LightGBM is used because the preceding N1 analysis identifies the nonlinear model with FEDE $\Delta$ as the strongest overall positively aligned response to the hidden channel.

The primary comparison is the true nonlinear direction $Y\rightarrow X$, evaluated by Eq.~\eqref{eq:directed_fede}. The reverse test exchanges source and target and evaluates $X\rightarrow Y$ with the same forecasting construction. Four independent candidate sources are additionally supplied to the $X$ forecast: (i) a sinusoid with an independent random phase, (ii) a quasiperiodic signal formed from two incommensurate sinusoidal components, (iii) a deterministic logistic-map signal at $r=4$ after a 500-sample burn-in, and (iv) an independent Gaussian AR(1) signal with coefficient 0.60, equal to the memory coefficient of the true source $Y$. Each control is standardized and generated independently for every realization, but is held fixed across the $\gamma$ and $\sigma$ scans within that realization. This design makes the AR(1) control particularly stringent: it matches the linear temporal memory of the true source while containing none of its imposed predictive contribution to $X$.

For every source/direction, $D_{\FEDE}^{v\rightarrow u}$ is averaged over the 30 stationary windows of each condition. A linear slope versus the complete $\gamma$ grid is then fitted separately for each realization, and the 20 slopes are summarized by $T$ and by a 95\% confidence interval for the mean slope. Because positive and negative control responses could otherwise cancel across noise levels, source specificity is summarized across $\sigma$ using mean $|T|$ in addition to the signed $T$ values. Benchmark N3 results are reported in Section~\ref{sec:res_n3_specificity}.

\subsection{Benchmark N4: functional-form method comparison}\label{sec:n4_method}
Benchmark N4 asks whether the response observed for N1 depends on the particular quadratic mapping $\phi_2$. The N1 quadratic channel in Eq.~\eqref{eq:phi} is retained as one member of the benchmark, and three additional normalized channels are introduced:
\begin{equation}
\begin{aligned}
\phi_{3}(y) &= \frac{y^3-3y}{\sqrt{6}},\\
\phi_{4}(y) &= \frac{y^4-6y^2+3}{\sqrt{24}},\\
\phi_{\sin}(y) &= \operatorname{standardize}\!\left[\sin(y^2)\right].
\end{aligned}
\label{eq:hidden_family}
\end{equation}
The three polynomial channels $\phi_2$, $\phi_3$, and $\phi_4$ are normalized probabilists' Hermite components for a Gaussian variable and are orthogonal to lower Gaussian orders in the population; $\phi_{\sin}$ is non-polynomial and even. For every retained realization, each channel is additionally centered and rescaled to exact sample mean zero and sample standard deviation one, so that $\gamma$ multiplies equal finite-record RMS channel power. Candidate realizations are rejected if $|r(y,q)|>0.15$ or if the finite-sample raw correlation $|r(y,\phi_k(y))|>0.05$ for any of the four channels. The same $y$ and $q$ trajectories, initial target state, and noise realization are reused across channel forms within a realization and noise level; only the hidden mapping changes. Predictors always receive raw $y_t$, never $\phi_k(y_t)$.

The remaining protocol preserves N1: 20 realizations, 3000 retained samples, $W=100$, step 100, history $m=1$, the 13-point coupling grid from 0 to 1, and
\begin{equation}
\sigma\in\{0,0.05,0.10,0.20\}.
\end{equation}
For FEDE and mean log-RMSE, LightGBM uses the same fixed hyperparameters as in N1 and produces matched forecasts for $h=1,\ldots,14$. Linear and degree-2 polynomial Granger-style scores use the chronological 60:40 out-of-sample protocol in Eq.~\eqref{eq:gc_oos}. CCM uses Eq.~\eqref{eq:ccm_delta}. Both the true $Y\rightarrow X$ and reverse $X\rightarrow Y$ directions are calculated.

For every method, channel, direction, noise level, and realization, the window-level scalar is averaged over the 30 non-overlapping windows. A realization-wise slope across the same 13-point $\gamma$ grid is then fitted and converted to the common external sensitivity statistic $T$ in Eq.~\eqref{eq:T}. Consequently, $T$ is not specific to FEDE: it is the common calibration scale used to compare how reproducibly each method responds to the known change in coupling. Signed $T$ is retained for all methods. Positive orientation is defined a priori as increasing predictive contribution with increasing $\gamma$; a negative $T$ therefore denotes a reproducible anti-aligned response. Benchmark N4 method-comparison and timing results are reported in Sections~\ref{sec:res_n4_methods} and~\ref{sec:res_n4_timing}.

\subsection{Benchmark N5: FEDE parameter robustness}\label{sec:n5_method}
The final synthetic benchmark tests whether the FEDE response depends on the nominal analysis scales used in the preceding experiments. The four hidden channels---$\phi_2$ from Eq.~\eqref{eq:phi} and the three alternatives in Eq.~\eqref{eq:hidden_family}---are evaluated at fixed $\sigma=0.10$, with 20 realizations, 3000 retained samples, the same 13-point coupling grid, the same LightGBM hyperparameters, and the true $Y\rightarrow X$ direction only. The predefined grid is
\begin{equation}
W\in\{100,200,400\},\qquad
m\in\{1,3,5\},\qquad
H\in\{5,14,20\},
\label{eq:param_grid}
\end{equation}
which gives 27 configurations per hidden channel and 108 FEDE $\Delta$ response estimates in total. Rolling step is set equal to $W$, so the local windows do not overlap. Each value of $H$ is fitted and evaluated independently rather than being truncated from a longer precomputed forecast profile. Within every configuration, the realization-wise FEDE slope across $\gamma$ is converted to $T$ using Eq.~\eqref{eq:T}. The original setting $W=100$, $m=1$, $H=14$ is retained as the primary reference and is not re-selected after inspecting the grid. Benchmark N5 results are reported in Section~\ref{sec:res_n5_robustness}.

\section{Results}
\subsection{Benchmark L1: calibration to the known coupling strength}\label{sec:res_expA_calibration}
\begin{figure}[H]
\centering
\includegraphics[width=\linewidth]{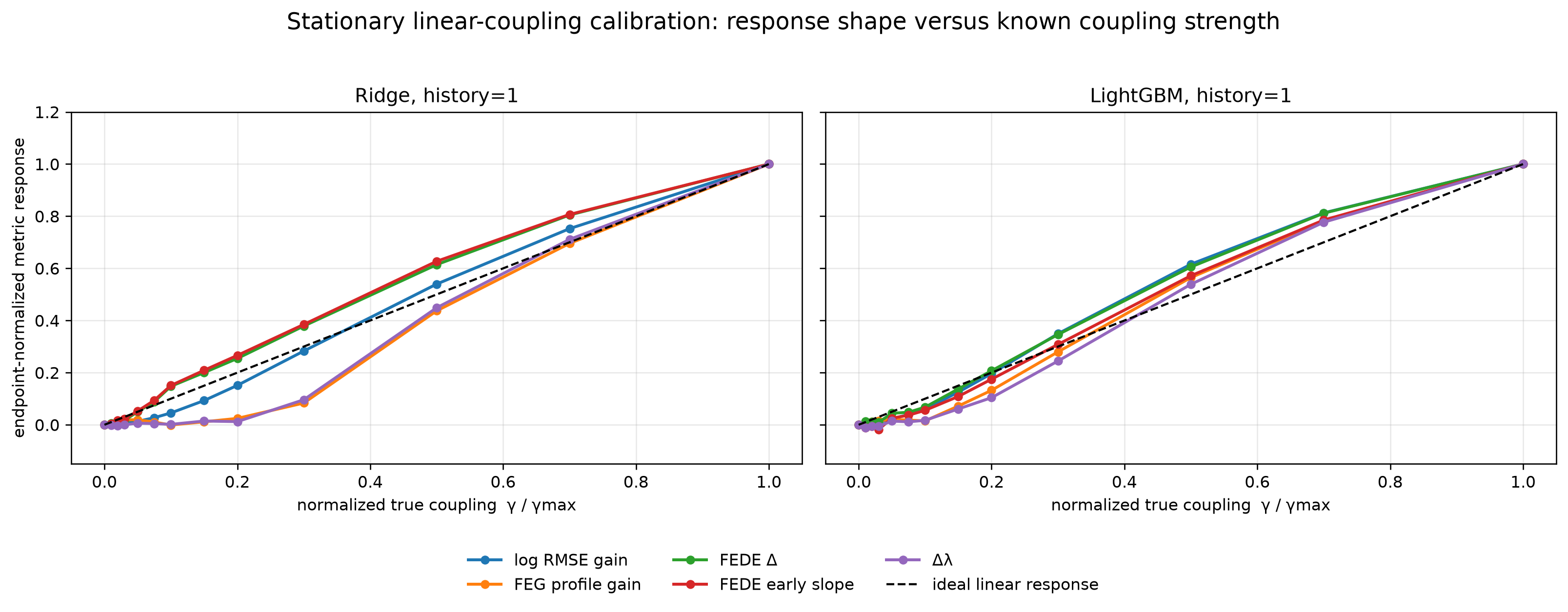}
\caption{Stationary linear-coupling calibration at $\sigma=0$. Each metric is endpoint-normalized to emphasize response shape rather than numerical scale. The dashed line is the ideal linear response.}
\label{fig:calibration}
\end{figure}

Benchmark L1 (Section~\ref{sec:expA_method}) provides the linear-reference calibration. Figure~\ref{fig:calibration} compares the normalized dose--response shapes at zero process noise. The dashed diagonal represents an ideal linear response to $\gamma$. Ridge provides a particularly clear linear reference. Among its FEG-based descriptors, FEDE delta and FEDE early-slope change follow the known coupling closely over most of the range, whereas FEG profile gain and $\Delta\lambda$ show stronger departures from linearity at weak coupling. LightGBM gives a smooth monotone response but is not expected to outperform a linear predictor in a system whose injected auxiliary term is itself linear.

The calibration $R^2$ values in Table~\ref{tab:linearity} make this comparison quantitative. At $\sigma=0$, Ridge FEDE delta and FEDE early-slope changes both achieve $R^2\approx0.99$. As noise increases, the Ridge early-slope descriptor remains close to linear over a broad range: $R^2=0.934$ at $\sigma=0.20$ and $0.862$ at $\sigma=0.80$. Conventional Ridge log-RMSE gain is slightly less linear at low noise but more stable at the highest noise level. Thus FEDE early slope is not a universal high-noise replacement for conventional error gain; its main advantage is the combination of low-noise calibration and weak-coupling sensitivity.

\begin{table}[H]
\centering
\caption{Linearity of the metric-versus-coupling calibration, measured by $R^2$ of a linear fit to the condition means. Selected history-$1$ metrics are shown.}
\label{tab:linearity}
\begin{tabular}{llccc}
\toprule
Predictor & Metric & $\sigma=0$ & $\sigma=0.20$ & $\sigma=0.80$ \\
\midrule
Ridge & mean log-RMSE gain & 0.952 & 0.941 & 0.920 \\
Ridge & FEDE delta change & 0.995 & 0.910 & 0.832 \\
Ridge & FEDE early-slope change & 0.994 & 0.934 & 0.862 \\
LightGBM & mean log-RMSE gain & 0.964 & 0.933 & 0.887 \\
LightGBM & FEDE delta change & 0.984 & 0.904 & 0.545 \\
LightGBM & FEDE early-slope change & 0.986 & 0.923 & 0.365 \\
\bottomrule
\end{tabular}
\end{table}

\subsection{Benchmark L2: weak-coupling response at zero process noise}\label{sec:res_expB_weak}
\begin{figure}[H]
\centering
\includegraphics[width=\linewidth]{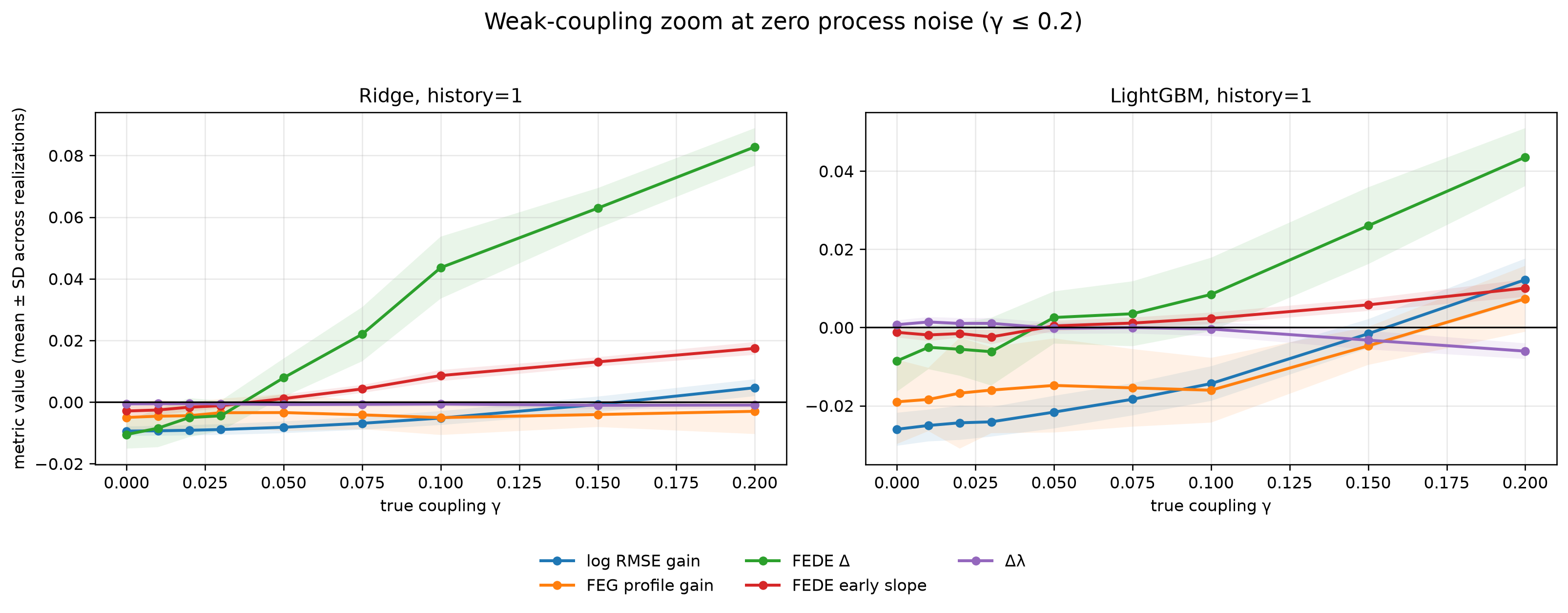}
\caption{Weak-coupling zoom at zero process noise. Lines show condition means and shaded bands show between-realization standard deviations. The Ridge FEDE descriptors respond to weak auxiliary coupling before a stable conventional log-RMSE gain appears.}
\label{fig:weakzoom}
\end{figure}

Benchmark L2 (Section~\ref{sec:expB_method}) targets weak linear coupling under process noise. Figure~\ref{fig:weakzoom} magnifies the regime $0\le\gamma\le0.20$. This view is central to the present study because a useful predictive-coupling descriptor should react before the auxiliary contribution becomes large. For Ridge, both FEDE delta and FEDE early-slope changes depart from their null levels at smaller $\gamma$ than the conventional mean log-RMSE gain. The FEDE delta response has the largest amplitude, while the early-slope response is smoother and remains close to a calibrated linear trend. By contrast, the Ridge FEG profile gain and $\Delta\lambda$ are weak in this low-coupling range.

\subsection{Benchmark L2: effect of process noise on calibration linearity}\label{sec:res_expB_noise}
\begin{figure}[H]
\centering
\includegraphics[width=\linewidth]{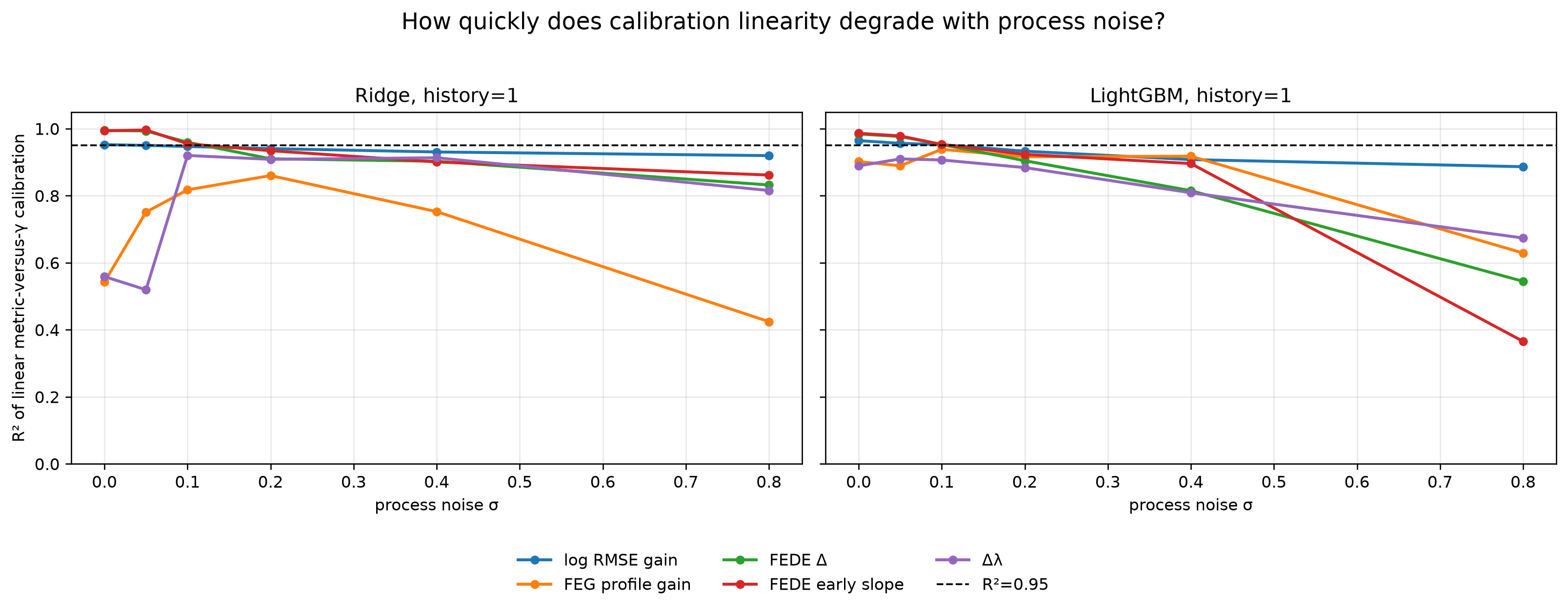}
\caption{Linearity of metric-versus-$\gamma$ calibration as process noise increases. The dashed line marks $R^2=0.95$.}
\label{fig:noise}
\end{figure}

Continuing benchmark L2, Fig.~\ref{fig:noise} shows how the linear calibration degrades as $\sigma$ increases. For Ridge, FEDE early-slope change stays near or above $R^2\approx0.9$ through $\sigma=0.40$, while FEG profile gain deteriorates much earlier. The conventional Ridge log-RMSE gain remains highly linear across the full noise grid. LightGBM FEDE responses lose calibration more rapidly at high noise, whereas its mean log-RMSE gain remains comparatively stable. These results support Ridge as the natural reference predictor for the present linear system and FEDE early-slope change as the preferred FEG-derived descriptor for subsequent comparison.

\subsection{Benchmark L2: minimum detectable predictive coupling}\label{sec:res_expB_detection}
\begin{figure}[H]
\centering
\includegraphics[width=\linewidth]{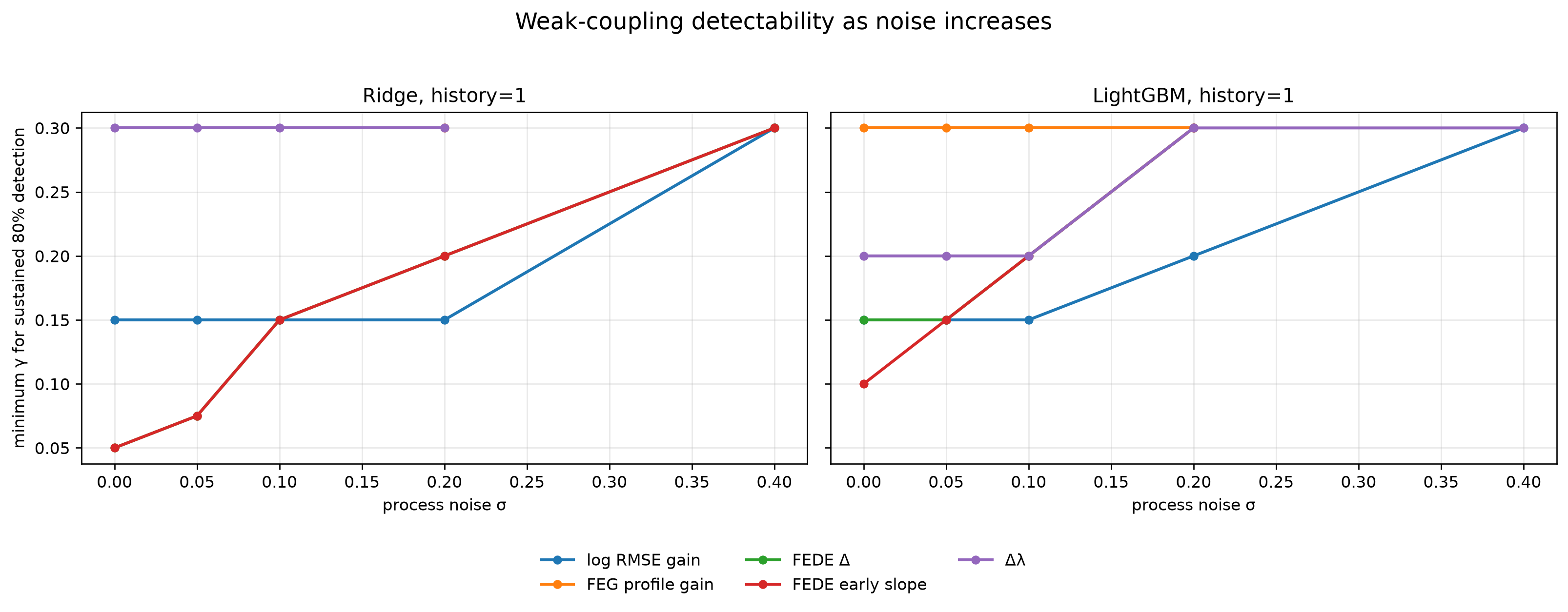}
\caption{Minimum coupling strength required for sustained 80\% null-relative detection using predictive metrics only. Missing points indicate that the criterion was not reached within the tested range $\gamma\le0.30$.}
\label{fig:detection}
\end{figure}

The weak-coupling experiment provides a second criterion: not just whether a metric is linear in $\gamma$, but how small a coupling can be separated from the exact $\gamma=0$ null. Figure~\ref{fig:detection} and Table~\ref{tab:gmin} summarize the sustained 80\% detection threshold.

At zero noise, Ridge mean log-RMSE gain requires $\gamma=0.15$, whereas both Ridge FEDE delta and FEDE early-slope change detect $\gamma=0.05$. At $\sigma=0.05$, the FEDE thresholds are $0.075$, still a factor of two below the conventional Ridge threshold of $0.15$. At $\sigma=0.10$ the difference disappears: all three reach $\gamma=0.15$. For larger noise, conventional log-RMSE gain is equal or more sensitive. LightGBM does not improve this linear benchmark; its FEDE early-slope threshold is $0.10$ at zero noise and $0.15$ at $\sigma=0.05$.

\begin{table}[H]
\centering
\caption{Sustained 80\% detection limit $\gmin$ for the main history-$1$ predictive metrics. $>0.30$ means that the criterion was not reached on the tested grid.}
\label{tab:gmin}
\begin{tabular}{llcccccc}
\toprule
Predictor & Metric & 0 & 0.05 & 0.10 & 0.20 & 0.40 & 0.80 \\
\midrule
Ridge & mean log-RMSE gain & 0.150 & 0.150 & 0.150 & 0.150 & 0.300 & $>0.30$ \\
Ridge & FEDE delta change & \textbf{0.050} & \textbf{0.075} & 0.150 & 0.200 & 0.300 & $>0.30$ \\
Ridge & FEDE early-slope change & \textbf{0.050} & \textbf{0.075} & 0.150 & 0.200 & 0.300 & $>0.30$ \\
LightGBM & mean log-RMSE gain & 0.150 & 0.150 & 0.150 & 0.200 & 0.300 & $>0.30$ \\
LightGBM & FEDE early-slope change & 0.100 & 0.150 & 0.200 & 0.300 & $>0.30$ & $>0.30$ \\
\bottomrule
\end{tabular}
\end{table}

The two result criteria therefore give a consistent picture. In the controlled linear regime, Ridge is the appropriate predictive reference. Within the Ridge error structure, FEDE early-slope change is a strong candidate descriptor because it is approximately proportional to coupling strength over a broad low-noise range and reaches the same favorable weak-coupling detection threshold as FEDE delta. FEDE delta remains an important companion descriptor because its low-coupling amplitude is larger. Neither descriptor should be interpreted as universally superior under strong process noise.

\subsection{Benchmark N1: hidden nonlinear channel validation}\label{sec:res_n1_validation}
\begin{figure}[H]
\centering
\includegraphics[width=0.98\linewidth]{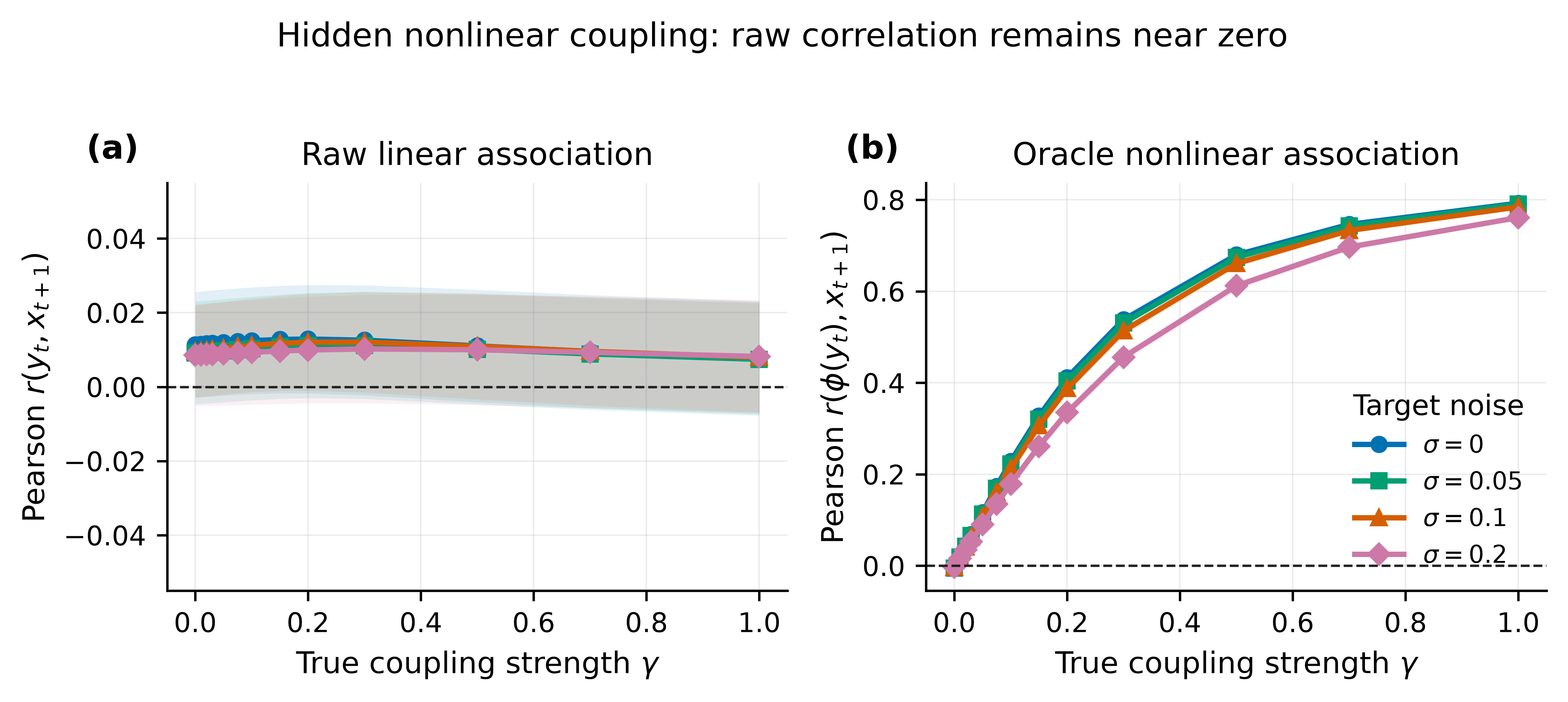}
\caption{Validation of the hidden nonlinear channel. Lines are means over 20 realizations and shaded bands are 95\% confidence intervals of the mean. (a) The raw one-step Pearson association between $y_t$ and $x_{t+1}$ remains near zero. (b) The synthetic oracle based on the known hidden transform $\phi(y_t)$ increases strongly with the imposed coupling $\gamma$.}
\label{fig:nonlinear_diag}
\end{figure}

Benchmark N1 (Section~\ref{sec:n1_method}) was designed so that a direct raw linear association cannot reveal the imposed auxiliary channel. Figure~\ref{fig:nonlinear_diag} verifies this construction. Across all tested $\gamma$, the mean one-step raw Pearson correlation $r(y_t,x_{t+1})$ remains close to zero. At the strongest coupling $\gamma=1$, the mean raw correlation is only 0.0079 at $\sigma=0$ and 0.0082 at $\sigma=0.20$. In contrast, the synthetic oracle $r(\phi_2(y_t),x_{t+1})$ rises monotonically with coupling and reaches 0.792 at $\sigma=0$, 0.784 at $\sigma=0.10$, and 0.761 at $\sigma=0.20$. Thus the predictive channel is strong in the generating coordinates while remaining almost invisible to ordinary linear association with raw $y$.

\subsection{Benchmark N1: nonlinear extraction---LightGBM versus Ridge}\label{sec:res_n1_extract}
\begin{figure}[H]
\centering
\includegraphics[width=\linewidth]{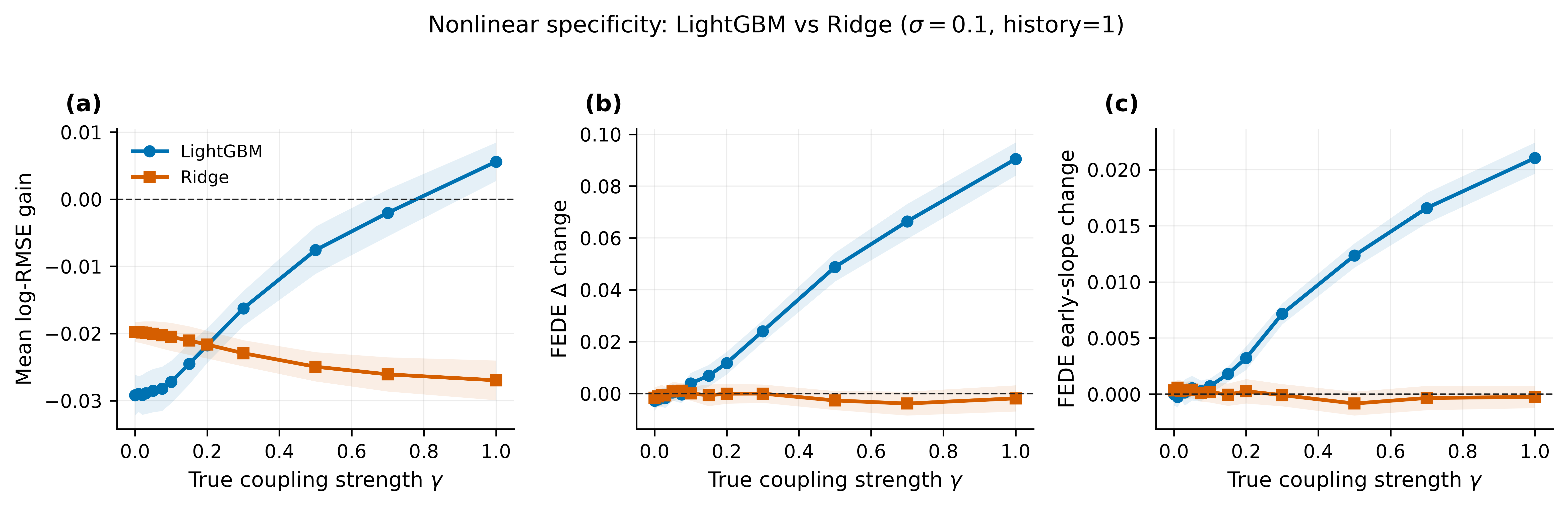}
\caption{Nonlinear specificity at $\sigma=0.10$ and history $m=1$. Each point is the mean of 20 realization-level values; each realization value is itself averaged over 30 non-overlapping $W=100$ windows. Shaded bands are 95\% confidence intervals of the mean across realizations. LightGBM develops a systematic coupling-dependent response, whereas Ridge remains close to its null FEDE response.}
\label{fig:nonlinear_specificity}
\end{figure}

Within benchmark N1 (Section~\ref{sec:n1_method}), Fig.~\ref{fig:nonlinear_specificity} compares the two predictor classes under moderate process noise, $\sigma=0.10$, using history $m=1$. This comparison is intentionally not made in the ideal noiseless case. The conventional mean log-RMSE gain improves gradually for LightGBM and becomes positive only at strong coupling, whereas the Ridge response becomes slightly more negative as $\gamma$ increases. The FEDE responses separate the predictor classes more clearly. LightGBM FEDE delta increases from $-0.0027$ at $\gamma=0$ to 0.0905 at $\gamma=1$, while Ridge remains near zero over the same range. LightGBM FEDE early-slope change similarly rises from approximately zero to 0.0210, while Ridge fluctuates around zero. Because neither predictor receives $\phi_2(y)$ explicitly, the difference isolates the ability of the nonlinear prediction engine to extract the hidden even dependence from raw $y$.

The response-sensitivity statistic $T$ makes the difference between metric amplitude and metric reproducibility explicit. Table~\ref{tab:T} reports $T$ for the three principal LightGBM descriptors over the complete stationary coupling grid. FEDE delta has the largest mean $T$ across the noise grid (26.71), closely followed by FEDE early-slope change (26.53); mean log-RMSE gain is substantially lower (14.92). At $\sigma=0.10$, the realization-wise mean slopes and 95\% confidence intervals are $0.0379\,[0.0321,0.0437]$ for log-RMSE gain, $0.0981\,[0.0902,0.1059]$ for FEDE delta, and $0.0230\,[0.0210,0.0250]$ for FEDE early slope. The corresponding $T$ values are 13.74, 26.24, and 24.24. Thus the two FEDE descriptors provide roughly twice the standardized dose-response strength of mean log-RMSE gain, despite having different numerical scales.

\begin{table}[H]
\centering
\scriptsize
\caption{Noise-normalized response sensitivity $T$ for LightGBM, history $m=1$, in the stationary hidden nonlinear benchmark. For each realization, a linear slope is fitted across the full 13-point $\gamma$ grid; $T$ is then computed from Eq.~\eqref{eq:T}. The second column gives the mean slope and its 95\% confidence interval at $\sigma=0.10$.}
\label{tab:T}
\begin{tabular}{lcccccc}
\toprule
Metric & $\bar\beta$ at $\sigma=0.10$ [95\% CI] & $T_{0}$ & $T_{0.05}$ & $T_{0.10}$ & $T_{0.20}$ & mean $T$ \\
\midrule
Mean log-RMSE gain & 0.0379 [0.0321, 0.0437] & 17.84 & 15.58 & 13.74 & 12.52 & 14.92 \\
FEDE $\Delta$ change & 0.0981 [0.0902, 0.1059] & 27.04 & 24.50 & \textbf{26.24} & \textbf{29.06} & \textbf{26.71} \\
FEDE early-slope change & 0.0230 [0.0210, 0.0250] & 25.87 & \textbf{29.80} & 24.24 & 26.21 & 26.53 \\
\bottomrule
\end{tabular}
\end{table}

The same calculation applied to Ridge provides a direct specificity control (Table~\ref{tab:T_ridge}). All three Ridge responses are negative across the tested noise grid. The FEDE values have small magnitudes, with mean $T=-0.77$ for FEDE delta and $T=-1.14$ for FEDE early slope, consistent with little reproducible monotone sensitivity to the hidden nonlinear channel. Mean log-RMSE gain shows a more systematic inverse response, with mean $T=-3.47$. At $\sigma=0.10$, its realization-wise mean slope is $-0.00815$ with 95\% CI $[-0.01253,-0.00377]$ and $T=-3.65$. Because positive log-RMSE gain is defined as improved prediction after adding $y$, this negative response means that the raw auxiliary variable increasingly \emph{worsens} the linear Ridge forecast as the strength of the hidden quadratic channel increases. It is therefore a reproducible anti-aligned response, not successful extraction of the nonlinear predictive contribution. The two Ridge FEDE slopes at $\sigma=0.10$ have confidence intervals spanning zero, in agreement with their $|T|$ values near one.

\begin{table}[H]
\centering
\scriptsize
\caption{Signed response sensitivity $T$ for Ridge, history $m=1$, in the stationary hidden nonlinear benchmark. Negative $T$ denotes a dose response opposite to the predefined positive orientation of the metric. A large $|T|$ would still indicate reproducible sensitivity, but in the inverse direction; values near zero indicate little reproducible response.}
\label{tab:T_ridge}
\begin{tabular}{lccccc}
\toprule
Metric & $T_{0}$ & $T_{0.05}$ & $T_{0.10}$ & $T_{0.20}$ & mean $T$ \\
\midrule
Mean log-RMSE gain & -3.37 & -3.56 & -3.65 & -3.32 & -3.47 \\
FEDE $\Delta$ change & -0.07 & -1.20 & -0.97 & -0.86 & -0.77 \\
FEDE early-slope change & -1.73 & -1.34 & -1.19 & -0.29 & -1.14 \\
\bottomrule
\end{tabular}
\end{table}

For comparison, the mean signed $T$ values of LightGBM are $14.92$, $26.71$, and $26.53$ for mean log-RMSE gain, FEDE delta, and FEDE early slope, respectively. Relative to Ridge, the nonlinear model therefore changes not only the magnitude but also the \emph{direction} of the calibrated response. This distinction is important when interpreting $T$: $|T|$ answers how reproducibly a metric changes with coupling, whereas the sign answers whether that change is aligned with the metric's predefined interpretation.

\begin{figure}[H]
\centering
\includegraphics[width=0.90\linewidth]{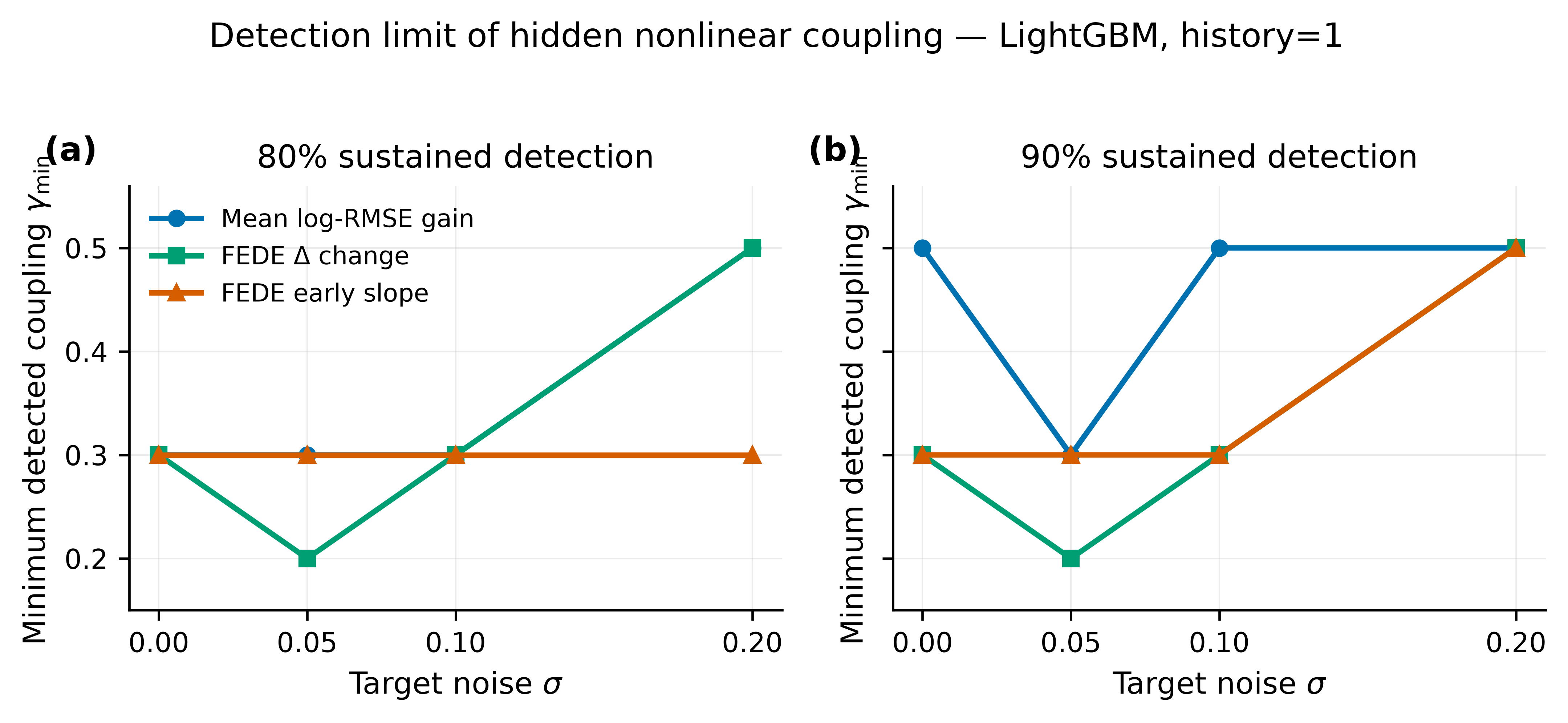}
\caption{Minimum hidden nonlinear coupling required for sustained 80\% and 90\% null-relative detection with LightGBM, history $m=1$. The FEDE descriptors reduce the strict 90\% threshold relative to mean log-RMSE gain in several low-to-moderate noise cases.}
\label{fig:nonlinear_detection}
\end{figure}

The null-relative detection analysis gives a complementary criterion. At $\sigma=0.10$, sustained 90\% detection is reached at $\gamma=0.30$ for both FEDE delta and FEDE early slope, compared with $\gamma=0.50$ for mean log-RMSE gain. At $\sigma=0$, the corresponding thresholds are 0.30, 0.30, and 0.50. At $\sigma=0.20$, all three require stronger coupling, with the 90\% criterion reached at $\gamma=0.50$. Figure~\ref{fig:nonlinear_detection} summarizes both the 80\% and 90\% criteria.

\subsection{Benchmark N3: directional and source specificity}\label{sec:res_n3_specificity}
The preceding stationary results establish that LightGBM can extract the hidden nonlinear channel and that FEDE $\Delta$ provides the largest overall noise-normalized response among the tested scalar descriptors. Benchmark N3 asks a stricter question: does the same response identify the correct candidate source and forecast direction, or does it also increase when an unrelated structured series is supplied to the augmented predictor?

The separation is pronounced (Table~\ref{tab:directional_specificity}). For the true $Y\rightarrow X$ channel, $T$ remains between 24.50 and 29.06 over all four noise levels, with mean $|T|=26.71$. Reversing the forecasting direction to $X\rightarrow Y$ reduces the response to $T=0.03$--1.23 and mean $|T|=0.60$. The independent periodic, quasiperiodic, chaotic, and AR(1) sources likewise remain close to the null response. Across all reverse and independent controls, the largest absolute value is $|T|=2.49$, observed for the sinusoidal control at $\sigma=0.20$. Consequently, even the smallest true-driver response, $T=24.50$, exceeds the largest individual control response by a factor of 9.8. On a noise-averaged absolute scale, the true-driver mean $|T|=26.71$ is approximately 24.8 times the largest control mean, 1.08 for the sinusoidal source.

\begin{table}[H]
\centering
\scriptsize
\caption{Directional and source specificity of the LightGBM FEDE $\Delta$ response. $T$ is the signed response-sensitivity statistic obtained from 20 realization-wise slopes across the complete coupling grid. The final column reports mean $|T|$ across noise levels so that positive and negative control responses cannot cancel.}
\label{tab:directional_specificity}
\begin{tabular}{lccccc}
\toprule
Candidate source / direction & $T_{0}$ & $T_{0.05}$ & $T_{0.10}$ & $T_{0.20}$ & mean $|T|$ \\
\midrule
$Y\rightarrow X$ (true nonlinear driver) & \textbf{27.04} & \textbf{24.50} & \textbf{26.24} & \textbf{29.06} & \textbf{26.71} \\
$X\rightarrow Y$ (reverse) & 0.03 & 0.99 & 1.23 & 0.15 & 0.60 \\
Sinusoid $\rightarrow X$ & 0.34 & 0.50 & 0.98 & 2.49 & 1.08 \\
Quasiperiodic $\rightarrow X$ & 1.01 & -0.07 & 0.26 & 0.59 & 0.48 \\
Logistic chaos $\rightarrow X$ & -0.55 & 1.21 & 0.42 & 0.29 & 0.62 \\
Independent AR(1) $\rightarrow X$ & 0.20 & -0.30 & 0.96 & 0.11 & 0.39 \\
\bottomrule
\end{tabular}
\end{table}

The realization-level slopes support the same interpretation. At $\sigma=0.10$, the true $Y\rightarrow X$ response has $\bar\beta=0.0981$ with 95\% CI $[0.0907,0.1054]$ and $T=26.24$. For the reverse $X\rightarrow Y$ test, the corresponding mean slope is only $\bar\beta=0.00255$ with 95\% CI $[-0.00151,0.00662]$ and $T=1.23$. The largest isolated control statistic, $T=2.49$ for the sinusoidal source at $\sigma=0.20$, is not reproduced across the remaining noise levels; moreover, its mean endpoint contrast between $\gamma=1$ and $\gamma=0$ is 0.00738 with 95\% CI $[-0.00047,0.01524]$, which spans zero. The control response is therefore consistent with a finite-sample fluctuation rather than a stable alternative predictive channel.

In benchmark N3 (Section~\ref{sec:n3_method}), the most stringent control is the independent AR(1) source because it has the same first-order memory coefficient, 0.60, as the true driver. Its mean $|T|$ is only 0.39. Likewise, the deterministic logistic and quasiperiodic controls do not acquire a systematic FEDE response as the true $Y\rightarrow X$ coupling increases. These results indicate that the FEDE $\Delta$ statistic is not merely reacting to added temporal structure, target complexity, or autocorrelation in the auxiliary input. Within this controlled benchmark it selectively tracks the candidate series that actually carries the imposed nonlinear predictive information, and it does so in the correct forecast direction.

\subsection{Benchmark N4: functional-form robustness and comparison with Granger and CCM}\label{sec:res_n4_methods}
Benchmark N4 (Section~\ref{sec:n4_method}) separates two different forms of performance: peak sensitivity when the comparator representation happens to match the hidden interaction, and robustness when the interaction form is unknown. Table~\ref{tab:method_comparison} reports the signed $T$ values averaged over the four noise levels for the true $Y\rightarrow X$ direction. The Granger entries are the chronological out-of-sample scores from Eq.~\eqref{eq:gc_oos}; the earlier in-window residual-ratio diagnostic is not used in the final comparison.

\begin{table}[H]
\centering
\scriptsize
\caption{Signed response-sensitivity statistic $T$, averaged over $\sigma\in\{0,0.05,0.10,0.20\}$, for four equal-power hidden nonlinear channels. Positive values are aligned with increasing useful predictive contribution. The polynomial Granger basis is fixed at degree two for all channels.}
\label{tab:method_comparison}
\begin{tabular}{lrrrr}
\toprule
Method & $H_2$ & $H_3$ & $H_4$ & $\sin(y^2)$ \\
\midrule
FEDE $\Delta$ + LightGBM & \textbf{29.00} & \textbf{27.47} & \textbf{23.22} & \textbf{28.67} \\
Mean log-RMSE + LightGBM & 14.09 & 10.68 & 12.67 & 12.71 \\
Linear Granger-style OOS & -7.94 & -3.08 & -6.70 & -3.56 \\
Polynomial Granger-style OOS, degree 2 & 74.21 & -6.77 & 3.53 & -4.01 \\
CCM convergence $\Delta\rho$ & 1.85 & 9.07 & 0.36 & 0.02 \\
\bottomrule
\end{tabular}
\end{table}

FEDE is the most uniform positively aligned response across the four nonlinear forms. Its noise-averaged $T$ varies only from 23.22 to 29.00, with mean $|T|=27.09$ across channel forms and a max/min ratio of 1.25. Even the smallest single-noise FEDE response in this benchmark is $T=20.29$ for $H_4$ at $\sigma=0.20$. By contrast, mean log-RMSE remains positive but is consistently weaker, with noise-averaged $T=10.68$--14.09. Because the two quantities are computed from the same LightGBM forecasts, this difference isolates the scalar summary: the entropy-based horizon structure provides a more reproducible coupling response than averaging forecast-error magnitude alone in this benchmark.

The out-of-sample Granger controls clarify the role of prior model form. Linear Granger-style gain is negative for all four channels, $T=-7.94$ to $-3.08$, meaning that adding raw $y_t$ increasingly worsens rather than improves held-out one-step prediction as the hidden nonlinear contribution grows. The degree-2 polynomial comparator behaves very differently when its prescribed basis matches the data-generating mechanism. For $H_2$, where the true channel is quadratic in $y$, it reaches $T=74.21$ and therefore exceeds FEDE by a wide margin. This is an expected advantage of a correctly specified low-dimensional parametric representation. However, the same fixed degree-2 model becomes anti-aligned for $H_3$ and $\sin(y^2)$ ($T=-6.77$ and $-4.01$) and only weakly positive for $H_4$ ($T=3.53$). Thus the benchmark does not support a claim that FEDE universally outperforms a matched parametric model; it supports the more specific conclusion that FEDE remains sensitive without being told which of several substantially different nonlinear forms generated the predictive channel.

The chronological scoring control is methodologically important. At $\gamma=0$, the mean OOS log gain is approximately $-0.03$ for the linear full model and $-0.11$ for the degree-2 full model, reflecting the ordinary generalization cost of extra parameters when the source contains no useful information. This behavior contrasts with the positive in-window nested residual ratios obtained when the same observations are used both to fit and score the larger model. The main comparison therefore uses only held-out Granger-style effect sizes.

Basic CCM convergence measures a different property and does not align consistently with the imposed predictive direction in this stochastic benchmark. Its true-direction noise-averaged $T$ is 1.85 for $H_2$, 9.07 for $H_3$, 0.36 for $H_4$, and 0.02 for $\sin(y^2)$. Moreover, the reverse-direction mean $|T|$ values in the same benchmark are 18.27, 18.34, 15.00, and 24.12, respectively. This should not be interpreted as a general failure of CCM. Here the source is stochastic AR(1), and three of the four mappings are even or otherwise non-invertible with respect to raw $y$, so the target can contain predictive information about $\phi_2(y)$ without preserving enough state information to reconstruct the signed source itself. The result instead emphasizes that forecast improvement and state-space reconstructability are distinct notions of directional dependence \cite{sugihara2012,monster2017}.

FEDE also preserves directionality across the functional-form scan. Its true/reverse mean-$|T|$ ratios are 30.2 for $H_2$, 67.6 for $H_3$, 45.3 for $H_4$, and 21.1 for $\sin(y^2)$. Thus the stable true-direction response is not obtained by making the descriptor indiscriminately responsive to either forecasting direction.

\subsection{Benchmark N2: intermittent hidden nonlinear coupling}\label{sec:res_n2_intermittent}
\begin{figure}[htbp]
\centering
\includegraphics[width=0.98\linewidth]{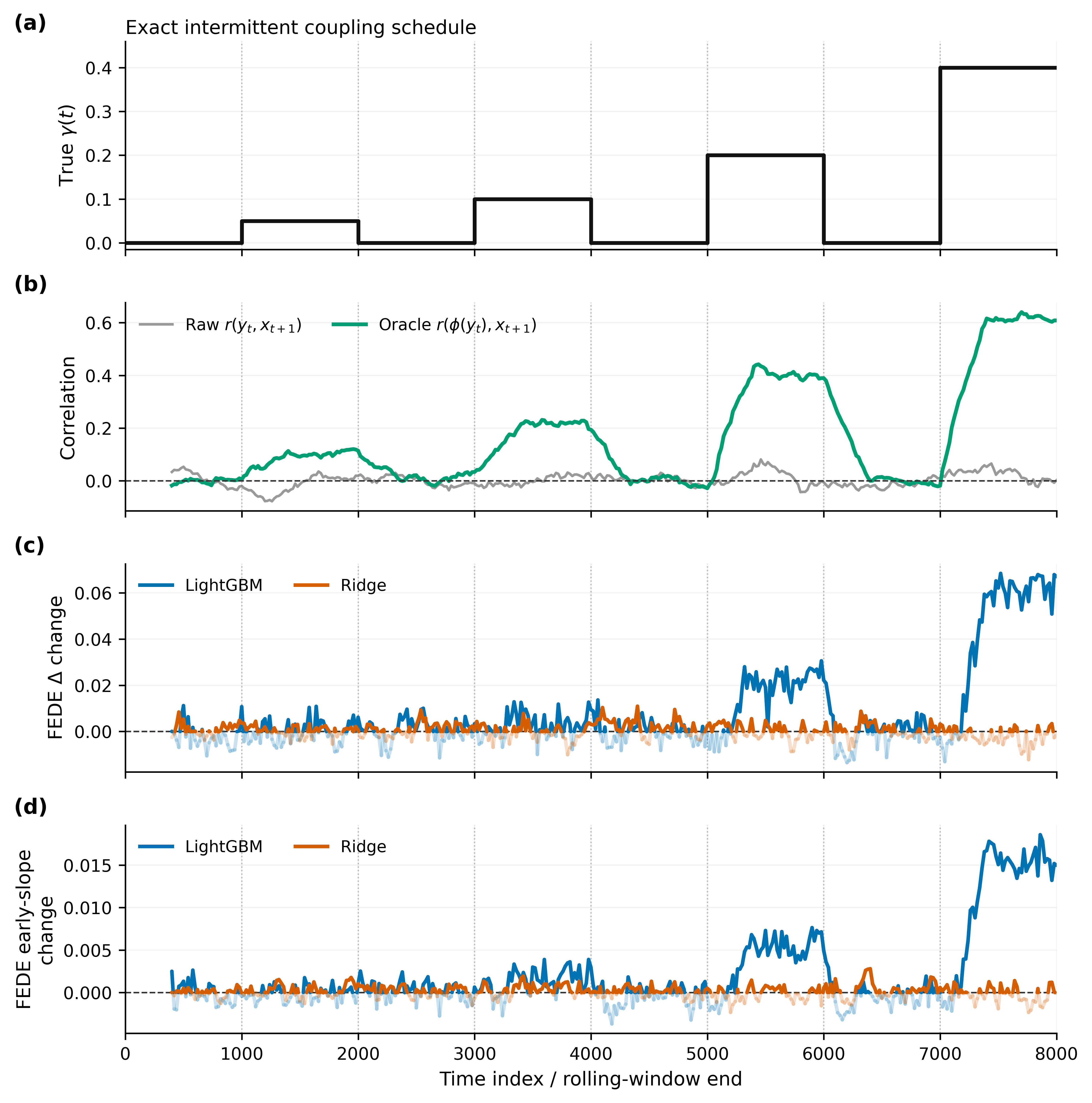}
\caption{Benchmark N2: time-resolved hidden nonlinear coupling at $\sigma=0.10$, history $m=1$, and rolling-window length $W=400$. (a) Exact imposed coupling schedule with 1000 samples per block. (b) Rolling raw linear correlation $r(y_t,x_{t+1})$ and nonlinear-oracle correlation $r(\phi_2(y_t),x_{t+1})$. (c) FEDE $\Delta$ change and (d) FEDE early-slope change for LightGBM and Ridge. Curves are unsmoothed ensemble means over 10 realizations with rolling step 20 and $H=14$. Negative FEDE values are retained numerically and shown with lower opacity rather than clipped.}
\label{fig:intermittent}
\end{figure}

Benchmark N2 (Section~\ref{sec:n2_method}) tests whether the nonlinear contribution can be followed in time rather than only calibrated across stationary realizations. Figure~\ref{fig:intermittent} shows the exact rectangular ground-truth schedule, the raw and oracle rolling correlations, and the two principal FEDE responses for the main-text setting $W=400$, $\sigma=0.10$, $m=1$, and rolling step 20. Each imposed coupling level lasts 1000 samples, so the active plateaus are long enough to separate the response level from the transition region even with the larger window. No post-hoc temporal smoothing is applied: the displayed curves are unsmoothed ensemble means over the 10 independent realizations.

The nonlinear oracle follows the active blocks and increases strongly with their amplitude, whereas the raw $r(y_t,x_{t+1})$ diagnostic remains close to zero. LightGBM FEDE responses are small at the weakest levels but form progressively clearer plateaus at $\gamma=0.20$ and $0.40$; Ridge remains near its baseline and does not reproduce this level ordering. The finite $W=400$ support visibly broadens the rising and falling edges relative to the exact rectangular schedule, which is the expected cost of the increased statistical averaging.

\begin{figure}[H]
\centering
\includegraphics[width=0.96\linewidth]{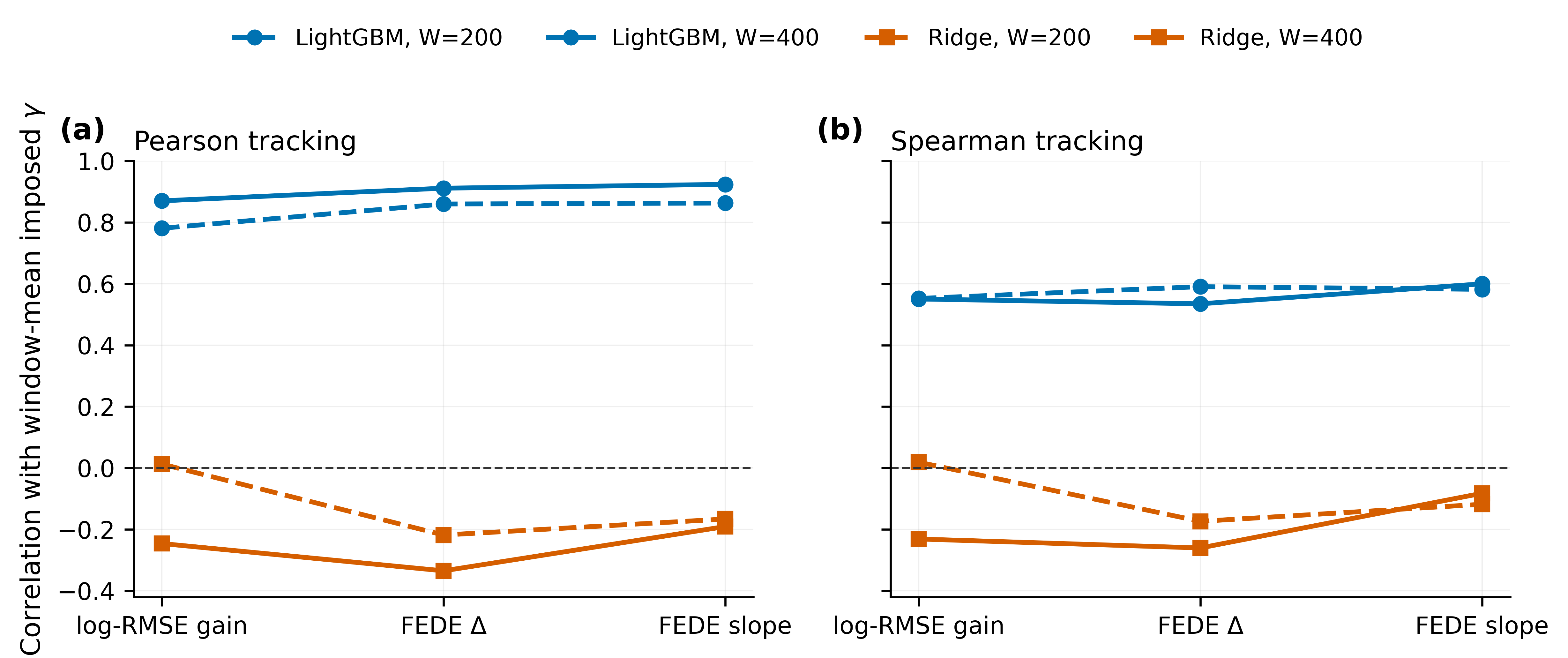}
\caption{Benchmark N2 tracking quality against the known window-mean imposed coupling $\bar\gamma(\tau)$. (a) Pearson and (b) Spearman correlations for mean log-RMSE gain, FEDE $\Delta$, and FEDE early-slope change. Solid curves use $W=400$ and dashed curves use $W=200$; circles denote LightGBM and squares denote Ridge. Correlations are computed from the unsmoothed ensemble-mean rolling responses.}
\label{fig:n2_tracking_quality}
\end{figure}

The quantitative tracking summary in Fig.~\ref{fig:n2_tracking_quality} uses the known window-mean imposed coupling $\bar\gamma(\tau)$ as the direct ground truth. For LightGBM at $W=400$, Pearson correlations are $r_P=0.870$ for mean log-RMSE gain, $0.911$ for FEDE $\Delta$, and $0.924$ for FEDE early slope; the corresponding Spearman values are $\rho_S=0.550$, $0.535$, and $0.600$. Against the nonlinear oracle, the same three LightGBM responses give Pearson correlations of $0.829$, $0.867$, and $0.883$, respectively, and Spearman correlations of $0.551$, $0.544$, and $0.598$. Thus FEDE early slope gives the strongest overall time-resolved tracking in this setting, with FEDE $\Delta$ close behind.

At $W=200$, the LightGBM Pearson correlations with $\bar\gamma(\tau)$ are $0.781$, $0.860$, and $0.863$ for log-RMSE gain, FEDE $\Delta$, and FEDE early slope, respectively; their Spearman counterparts are $0.552$, $0.590$, and $0.582$. Increasing the window to 400 therefore strengthens the linear alignment and stabilizes the visible plateau response, while the rank correlations remain of similar magnitude and the transition edges become broader. Ridge remains weak or anti-aligned at $W=400$: its Pearson correlations with $\bar\gamma(\tau)$ are $-0.246$, $-0.335$, and $-0.191$, and its corresponding Spearman correlations are $-0.231$, $-0.261$, and $-0.082$.

The $W=200$ time-domain curves are retained in Appendix~\ref{sec:appendix_n2} as a direct higher-resolution comparison. They show the complementary side of the trade-off: shorter windows localize switching more sharply but produce a visibly noisier response. The main-text $W=400$ choice is therefore a presentation setting for the intermittent benchmark, not a universal recommendation for empirical data.

\subsection{Benchmark N4: computational cost}\label{sec:res_n4_timing}
Benchmark N4 (Section~\ref{sec:n4_method}) was executed under Windows 11 with 16 logical CPUs, Python 3.14.5, scikit-learn 1.9.0, LightGBM 4.7.0, and 10 outer worker processes. The complete Program08 run, including four channels, four noise levels, 13 coupling levels, 20 realizations, true and reverse directions, LightGBM forecasting, Granger diagnostics, and CCM, required 104.6 minutes of wall time.

Table~\ref{tab:timing_methods} separates the principal true-direction method costs per 100 window evaluations. FEDE and mean log-RMSE share exactly the same LightGBM forecasts and therefore should not be added as independent costs. The shared LightGBM forecasting core requires 28.81 s per 100 windows, whereas the complete FEDE calculation requires 29.45 s. Thus the entropy-profile calculation itself adds approximately 0.64 s per 100 windows, or 2.2\% beyond the nonlinear forecasting core. The chronological OOS Granger controls are orders of magnitude cheaper, and basic CCM lies between OLS and LightGBM.

\begin{table}[H]
\centering
\scriptsize
\caption{Approximate computation time for 100 true-direction window evaluations. FEDE and mean log-RMSE share the same LightGBM forecast core. OOS Granger timings come from the dedicated chronological control run.}
\label{tab:timing_methods}
\begin{tabular}{lrr}
\toprule
Method & seconds / 100 windows & relative to FEDE \\
\midrule
FEDE $\Delta$ + LightGBM & 29.451 & 1.000 \\
Mean log-RMSE + LightGBM & 28.820 & 0.979 \\
CCM convergence $\Delta\rho$ & 1.864 & 0.063 \\
Linear Granger-style OOS & 0.0158 & 0.00054 \\
Polynomial Granger-style OOS, degree 2 & 0.0207 & 0.00070 \\
\bottomrule
\end{tabular}
\end{table}

The practical implication is that FEDE is not computationally cheap if a nonlinear forecast engine must first be fitted across many horizons. However, once those forecasts already exist, the entropy descriptor adds little additional cost. This distinction is relevant when FEDE is used as an extra diagnostic alongside an existing multi-horizon forecasting pipeline.

\subsection{Benchmark N5: parameter robustness of the FEDE response}\label{sec:res_n5_robustness}
Benchmark N5 (Section~\ref{sec:n5_method}) asks whether the positive FEDE dose response survives changes in the method's own analysis scales. At $\sigma=0.10$, the complete grid in Eq.~\eqref{eq:param_grid} contains 27 configurations for each of the four hidden nonlinear channels, giving 108 independently fitted FEDE $\Delta$ response statistics. Every value is positively aligned with the imposed coupling: all 108 values satisfy $T>0$, the smallest observed value is $T=16.49$, and the 95\% confidence interval of the realization-level coupling slope remains above zero for every tested grid point.

The original configuration used throughout the main nonlinear experiments, $W=100$, $m=1$, and $H=14$, remains strongly responsive without any post-hoc retuning. It gives $T=32.95$, 24.31, 20.35, and 29.71 for $H_2$, $H_3$, $H_4$, and $\sin(y^2)$, respectively. Across the full 27-point grid for each channel, the corresponding mean $T$ values are 41.34, 28.32, 22.30, and 45.39, while even the channel-specific minima remain between 16.49 and 18.02 (Table~\ref{tab:param_robustness}). Thus the positive response is not confined to the nominal parameter triplet.

\begin{table}[H]
\centering
\small
\caption{FEDE parameter robustness at $\sigma=0.10$. ``Primary'' is the configuration $W=100$, $m=1$, $H=14$ used in the main nonlinear experiments. Grid statistics summarize all 27 combinations in Eq.~\eqref{eq:param_grid} for each hidden channel.}
\label{tab:param_robustness}
\begin{tabular}{lrrrr}
\toprule
Hidden channel & Primary $T$ & Grid mean $T$ & Grid min $T$ & Grid max $T$ \\
\midrule
$H_2$ & 32.95 & 41.34 & 16.49 & 55.35 \\
$H_3$ & 24.31 & 28.32 & 16.90 & 42.88 \\
$H_4$ & 20.35 & 22.30 & 18.02 & 32.13 \\
$\sin(y^2)$ & 29.71 & 45.39 & 17.01 & 73.78 \\
\bottomrule
\end{tabular}
\end{table}

\begin{figure}[H]
\centering
\begin{minipage}[t]{0.49\linewidth}
\centering
\textbf{(a)}\\[-0.3em]
\includegraphics[width=\linewidth]{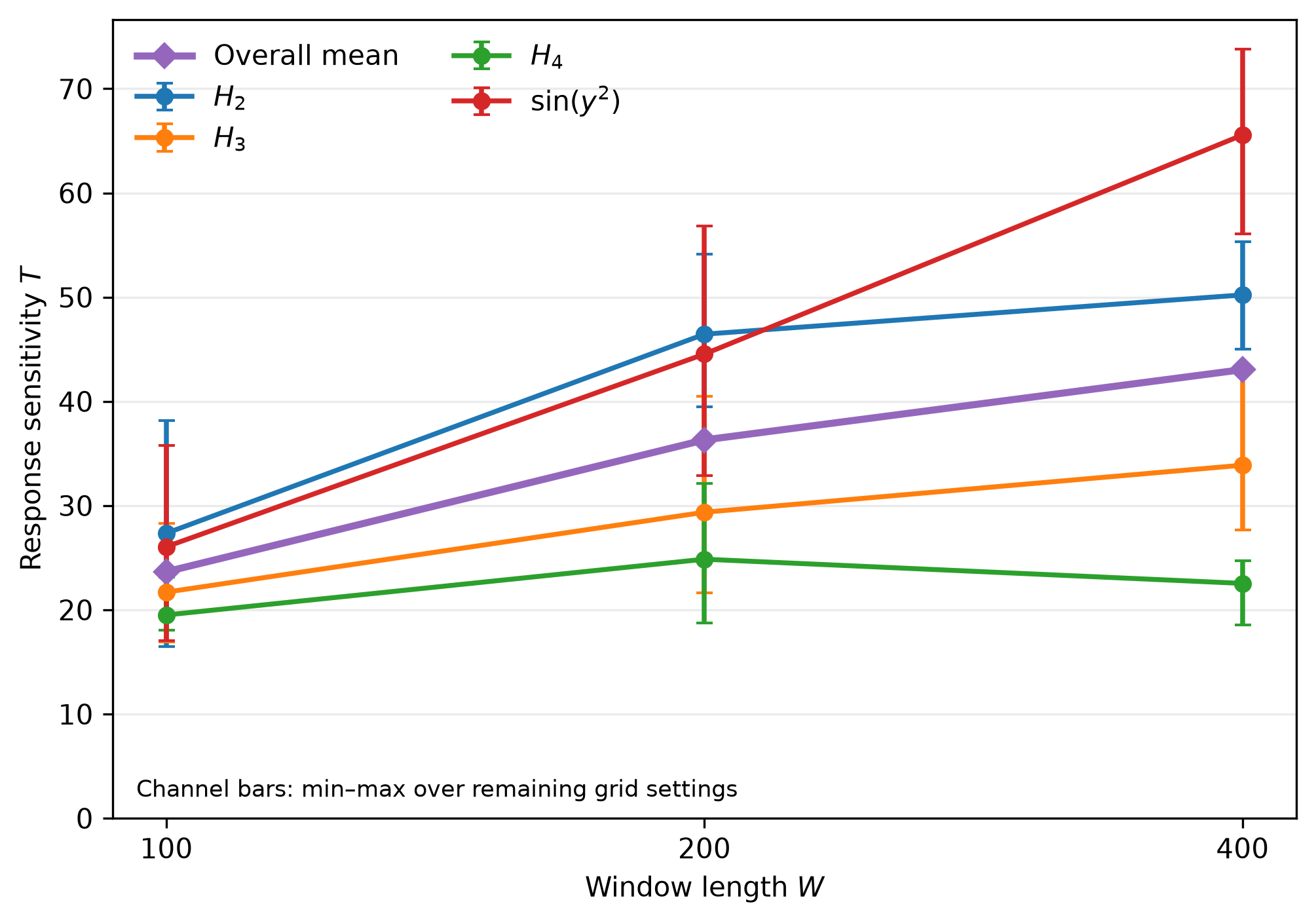}
\end{minipage}\hfill
\begin{minipage}[t]{0.49\linewidth}
\centering
\textbf{(b)}\\[-0.3em]
\includegraphics[width=\linewidth]{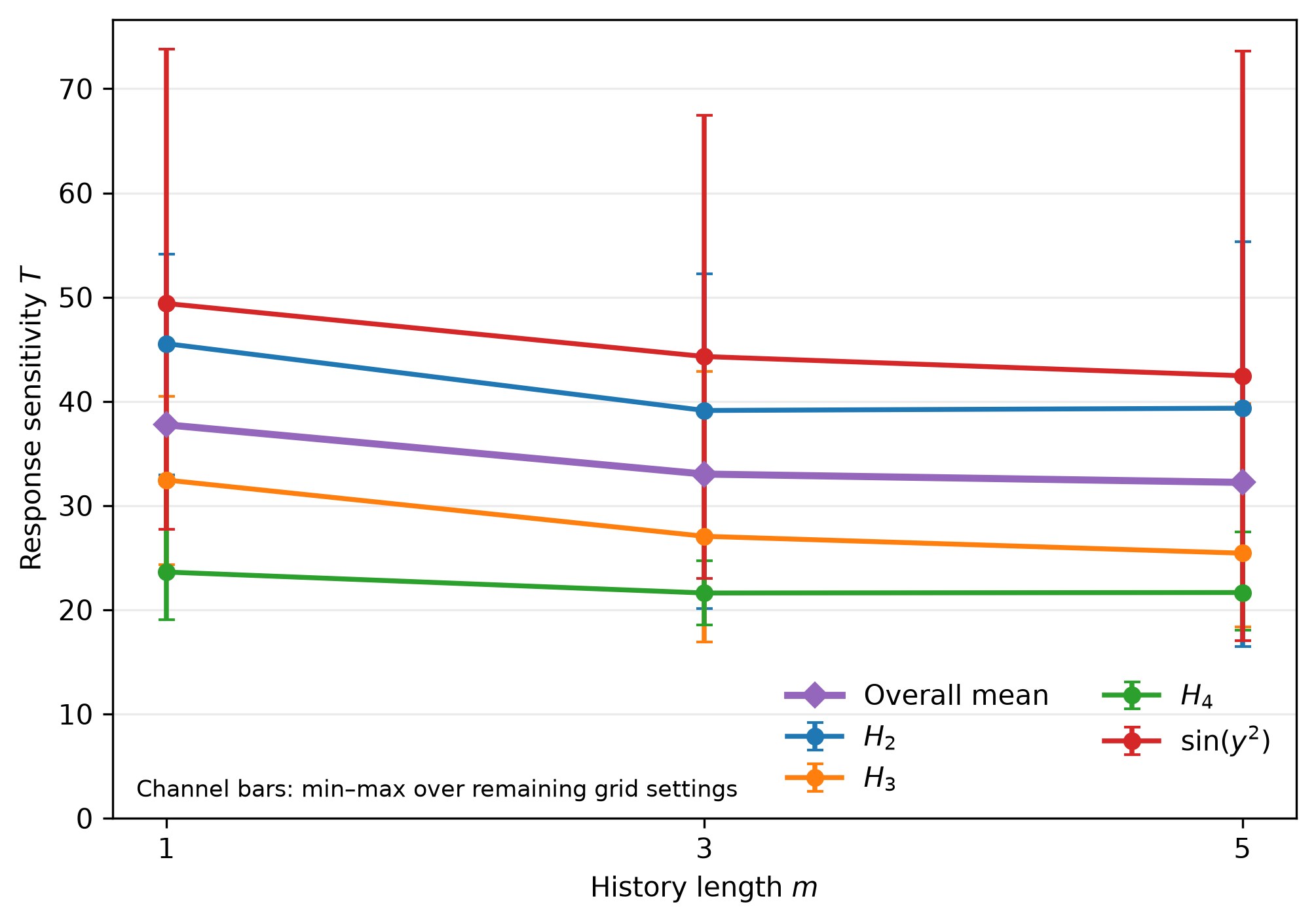}
\end{minipage}

\vspace{0.5em}
\begin{minipage}[t]{0.49\linewidth}
\centering
\textbf{(c)}\\[-0.3em]
\includegraphics[width=\linewidth]{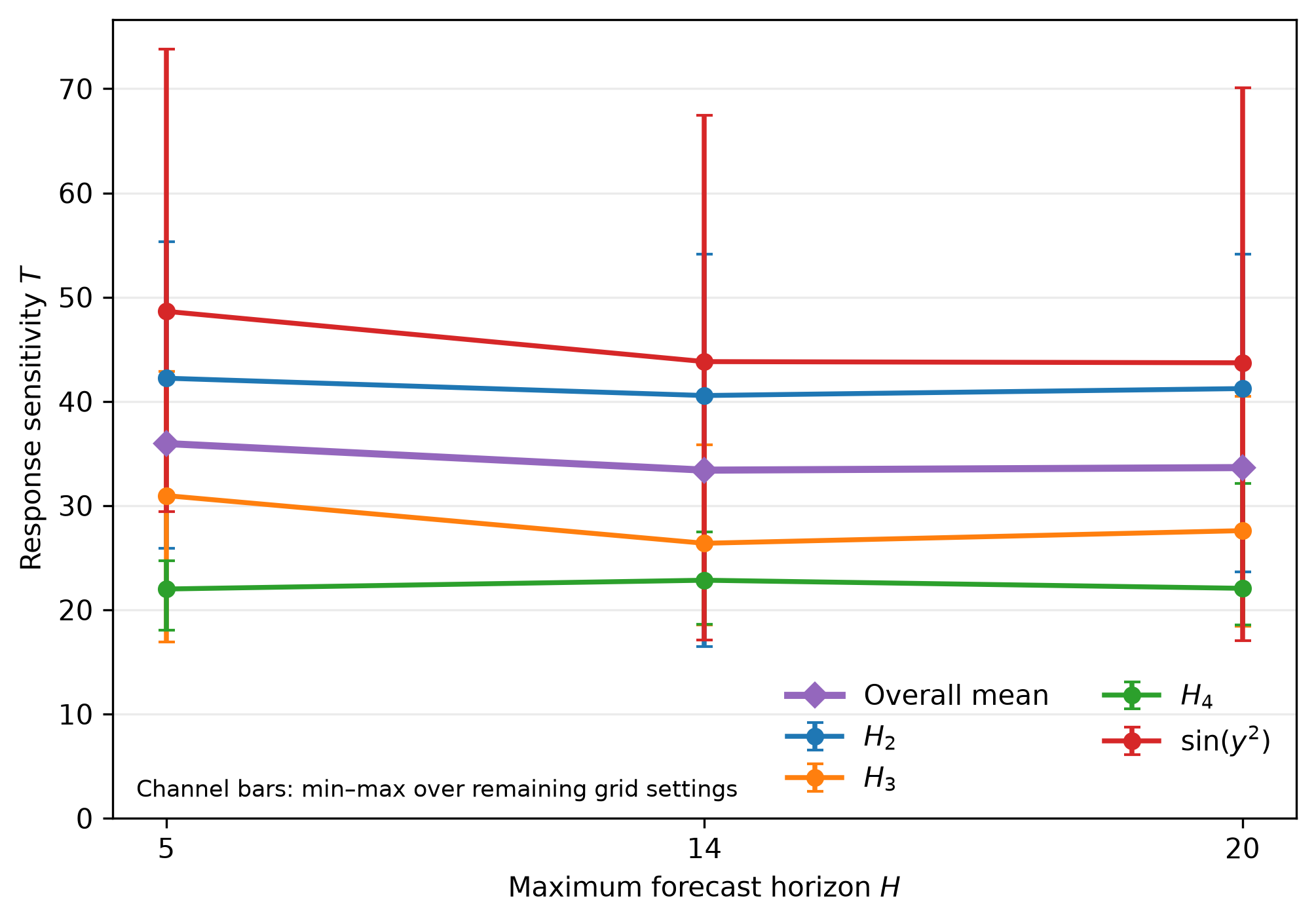}
\end{minipage}
\caption{Marginal FEDE response sensitivity over the predefined parameter grid at $\sigma=0.10$. (a) Window length $W$; (b) history length $m$; (c) maximum forecast horizon $H$. Curves show the four hidden nonlinear channels and their overall mean. For each channel, the marker is the mean $T$ after averaging over the other two grid parameters; vertical bars span the minimum and maximum over those remaining configurations and are not confidence intervals.}
\label{fig:param_effects}
\end{figure}

Figure~\ref{fig:param_effects} separates the marginal effects of the three analysis parameters. Window length produces the clearest aggregate change: after averaging over hidden channel, history, and horizon, mean $T$ rises from 23.65 at $W=100$ to 36.30 at $W=200$ and 43.05 at $W=400$. The corresponding mean realization-level slope increases from 0.0836 to 0.1222 and 0.1406. History length has a smaller systematic effect, with mean $T=37.76$, 33.03, and 32.23 for $m=1$, 3, and 5. Dependence on the maximum forecast horizon is weaker still over the tested range: mean $T=35.96$, 33.40, and 33.65 for $H=5$, 14, and 20, respectively. In the panels, each channel point is the marginal mean over the two remaining grid dimensions and its vertical bar is the corresponding min--max range, not a confidence interval.

The most robust single grid point under a worst-channel criterion is $W=200$, $m=1$, $H=20$, for which the weakest of the four channel responses is still $T=32.13$ and the mean across channels is 43.77. This point is reported descriptively rather than adopted as a replacement for the original setting. The sensitivity experiment is therefore interpreted as a robustness check over a predefined grid, not as a post-hoc search for a new optimum.

\section{Discussion}

\subsection{FEDE as a horizon-resolved predictive observable}
The central methodological result of this study is not that restricted-versus-augmented forecasting is new. That logic is already fundamental to Granger-style analysis, out-of-sample forecast comparison, and several nonlinear predictive frameworks \cite{shojaie2021,tank2021,rosol2022,diebold1995}. The distinct object introduced here is the \emph{horizon-resolved entropy evolution of the signed forecast-error distribution}. We therefore use the term \emph{FEDE predictive-causality framework} for the complete matched construction, with predictive causality understood only in the Granger sense of additional predictive information and not as structural or physical causal identification. FEDE sits downstream of a predictor: it asks whether adding a candidate source changes how predictive uncertainty is organized from short to longer horizons, rather than only whether a one-step residual variance or a horizon-averaged loss decreases. This distinction is summarized conceptually in Table~\ref{tab:intro_methods} and operationally in Table~\ref{tab:method_objects}.

The controlled experiments show why this extra structure can be informative. In the Gaussian-AR(1) nonlinear benchmark family N1--N5, raw $r(y_t,x_{t+1})$ remains close to zero even as the oracle association with the hidden transform increases (Fig.~\ref{fig:nonlinear_diag}). This behavior is deliberate: unlike the logistic-map linear references L1--L2, the nonlinear family injects the source through a transform that is orthogonal to raw Gaussian $y$ at first order. A nonlinear forecasting engine can nevertheless recover the predictive contribution, and FEDE converts that recovery into a substantially more reproducible dose response than the mean forecast-error magnitude. With LightGBM, the FEDE $\Delta$ and early-slope statistics reach mean signed sensitivities of 26.71 and 26.53, respectively, compared with 14.92 for mean log-RMSE gain (Table~\ref{tab:T}); the corresponding Ridge responses are weak or anti-aligned (Table~\ref{tab:T_ridge} and Fig.~\ref{fig:nonlinear_specificity}). Thus FEDE does not create nonlinear information that is absent from the predictor. Rather, once a predictor extracts that information, the entropy profile provides a different and, in these benchmarks, more stable summary of how the added source reshapes the forecast-error field across lead times.

This distinction is also visible in the detection experiment. At $\sigma=0.10$, sustained 90\% detection occurs at $\gamma=0.30$ for both FEDE descriptors but only at $\gamma=0.50$ for mean log-RMSE gain (Fig.~\ref{fig:nonlinear_detection}). Because all three descriptors are computed from the same LightGBM forecasts, this improvement cannot be attributed to a stronger prediction model. It reflects the choice of response variable itself: the distributional reorganization of errors across horizons can be more reproducible than a scalar average of their magnitudes.

Benchmark N2 extends this observation from stationary dose--response curves to a time-varying setting. With 1000-sample coupling blocks, the $W=400$ LightGBM FEDE early-slope response reaches $r_P=0.924$ with the known window-mean coupling and $r_P=0.883$ with the nonlinear oracle, while FEDE $\Delta$ reaches 0.911 and 0.867, respectively. The shorter $W=200$ analysis retains sharper transition localization but lower Pearson alignment. This is not a contradiction: increasing $W$ supplies more forecast errors for entropy estimation and stabilizes the response, but any rolling statistic necessarily averages over a wider temporal support. The intermittent benchmark therefore turns the window-size effect into an explicit resolution--stability trade-off rather than a tuning artifact.

\subsection{Comparison with Granger-style and conventional forecast-gain approaches}
The method-comparison benchmark in Table~\ref{tab:method_comparison} provides the most direct context for interpreting FEDE relative to established predictive approaches. The literature already offers a continuum from classical linear Granger causality to polynomial/Volterra, kernel, neural, and other machine-learning extensions \cite{shojaie2021,tank2021,rosol2022,hasanzadeh2021,wismuller2021}. Their common strength is a clear restricted-versus-full interpretation; their common dependence is that sensitivity is tied to the representation supplied to the model. Our results reproduce both sides of that trade-off.

A correctly specified parametric comparator can outperform FEDE by a large margin. For the quadratic $H_2$ channel, the fixed degree-2 polynomial Granger-style model explicitly contains the relevant $y_t^2$ term and reaches $T=74.21$, whereas FEDE reaches $T=29.00$. This is not a weakness to be hidden: when the interaction form is known a priori, a parsimonious matched model is preferable and can be far more sensitive. The more revealing result appears when the hidden mapping changes while the polynomial basis is left fixed. The same degree-2 comparator becomes anti-aligned for $H_3$ and $\sin(y^2)$ and only weakly positive for $H_4$, while FEDE remains positively aligned in the narrow range $T=23.22$--29.00 across all four nonlinear forms. The advantage demonstrated here is therefore \emph{robustness to unknown interaction form}, not universal superiority over a model supplied with the correct basis.

The linear Granger-style control further illustrates the importance of honest out-of-sample scoring. Using a chronological 60:40 split, the linear full model is anti-aligned for every hidden nonlinear channel ($T=-7.94$ to $-3.08$), and the degree-2 model is strongly useful only in its matched quadratic case. In contrast, an in-window nested residual comparison gives the larger model an automatic training-error advantage. The OOS formulation in Eq.~\eqref{eq:gc_oos} removes that bias and better matches the operational question posed in this paper: does the candidate source improve unseen future prediction? This choice is consistent with the broader forecast-comparison literature, where predictive value is evaluated on held-out errors and formal procedures such as Diebold--Mariano tests address whether forecast-loss differences are statistically distinguishable from zero \cite{diebold1995,grant2026}. FEDE is complementary to such tests: it supplies a magnitude/structure descriptor, whereas a DM-type procedure supplies inferential evidence about a chosen loss differential.

The comparison with mean log-RMSE is especially informative because the prediction engine is held fixed. Across $H_2$, $H_3$, $H_4$, and standardized $\sin(y^2)$, FEDE yields noise-averaged $T=29.00$, 27.47, 23.22, and 28.67, whereas the corresponding mean log-RMSE values are 14.09, 10.68, 12.67, and 12.71 (Table~\ref{tab:method_comparison}). FEDE therefore does more than re-express forecast accuracy on a different scale. It uses the \emph{shape change} of the signed-error distribution across $h=1,\ldots,H$, which appears to preserve a stronger and more uniform coupling signature than the mean error magnitude in the present controlled setting.

A possible explanation for this sensitivity is that RMSE compresses each horizon-specific error distribution to its second moment, whereas FEDE responds to changes in how probability mass is distributed across the error range. An auxiliary source may therefore reorganize the forecast errors---by changing their concentration, tails, or other non-Gaussian distributional features---without producing a proportionally large change in mean squared magnitude. FEDE additionally tracks how this reorganization evolves with forecast horizon, allowing a weak but systematic distributional signature to accumulate into a stable horizon-resolved response. Thus, the observed gain arises not from changing the forecasts, but from changing what is measured in their errors. Importantly, no qualitatively new information should be expected if the only effect were a pure Gaussian scale change: for an ideal Gaussian scale family, differential entropy is a monotonic function of variance. The stronger FEDE response observed here therefore suggests that the nonlinear predictive channel modifies the structure and horizon evolution of the forecast-error distribution rather than merely rescaling its variance. This interpretation is qualitative rather than a formal equivalence for the finite-bin entropy estimator used in the present implementation.

\subsection{Relation to CCM and state-space directionality}
CCM and EDM occupy a different conceptual position in the literature. CCM asks whether information about a putative driver is recoverable from the reconstructed manifold of the affected variable and whether this cross-map skill converges with increasing library size \cite{sugihara2012,chang2017}. EDM and S-map methods extend state-space reconstruction toward nonlinear forecasting and state-dependent interaction strengths \cite{deyle2016,chang2017}. These are powerful tools, but they do not estimate the same object as incremental forecast contribution, and comparative reviews emphasize that no single directional measure is uniformly optimal across deterministic, stochastic, noisy, or synchronized regimes \cite{edinburgh2021}.

That distinction is central to the present CCM result. Basic CCM convergence is weak or directionally inconsistent for the stochastic AR(1) source and partly non-invertible hidden mappings in Table~\ref{tab:method_comparison}. For $H_2$, $H_4$, and $\sin(y^2)$, opposite signs of raw $Y$ can produce the same hidden contribution, so knowledge of $Y$ can improve prediction of $X$ even though the state of $X$ cannot uniquely reconstruct the signed source. Process noise and external influence are also known to complicate cross-mapping behavior \cite{monster2017}. The observed weak or reversed CCM response should therefore be read as a scope result, not as evidence that CCM is inferior in the deterministic attractor regimes for which it was developed.

The directional/source-specificity experiment addresses a different possible criticism: that FEDE might respond simply because \emph{any} structured auxiliary variable is added to the model. Table~\ref{tab:directional_specificity} argues against that interpretation. The true $Y\rightarrow X$ channel gives mean $|T|=26.71$, while reverse direction and independent periodic, quasiperiodic, chaotic, and AR(1) controls remain at mean $|T|\le1.08$; even the smallest true-driver response exceeds the largest individual control response by a factor of 9.8. The four-channel experiment then preserves large true/reverse separation under substantially different nonlinear forms. These results support \emph{directional predictive specificity} within the controlled benchmark, while stopping short of structural causal identification. Common drivers, hidden confounders, feedback, and incomplete conditioning can still make predictive direction differ from physical causation, as is well recognized in both Granger and state-space analyses \cite{shojaie2021,stokes2017,monster2017}.

\subsection{Computational cost and practical trade-offs}
The computational comparison in Table~\ref{tab:timing_methods} makes clear that FEDE is not the fastest method considered. For 100 true-direction window evaluations, the full FEDE+LightGBM pipeline requires approximately 29.45~s, compared with 1.86~s for the present basic CCM implementation and only 0.0158--0.0207~s for the OOS linear and degree-2 polynomial Granger controls. If the only objective is a one-step restricted-versus-full test under a plausible low-dimensional parametric model, the OLS-based approaches are decisively cheaper.

The practical comparison changes, however, when multi-horizon nonlinear forecasts already form part of the analysis. Mean log-RMSE and FEDE share the same LightGBM forecast core, which costs approximately 28.81~s per 100 windows. The complete FEDE calculation costs 29.45~s, so the entropy-profile processing itself adds only about 0.64~s per 100 windows, or 2.2\% beyond the shared forecast core. FEDE is therefore best viewed not as a computational replacement for Granger or CCM, but as a low-overhead \emph{additional diagnostic} once a flexible multi-horizon forecast pipeline is already being run. The cost is paid mainly for the nonlinear forecasts; the distributional descriptor extracts extra information from forecast errors that have already been generated.

This trade-off also clarifies when FEDE is most attractive. A simple linear or polynomial model is preferable when its assumptions are credible and computational economy is paramount. A state-space method is more natural when manifold reconstruction and deterministic interaction geometry are the scientific object. FEDE becomes attractive when the forecasting system is already nonlinear, the functional form of the candidate interaction is not known in advance, and one wants a directional descriptor that uses the full horizon evolution rather than only a one-step loss or a single averaged accuracy score. This positioning is consistent with the method landscape summarized in Table~\ref{tab:intro_methods}: the methods are complementary because they answer related but non-identical questions.

\subsection{Parameter robustness and the resolution--stability trade-off}
The parameter-grid experiment closes an important internal robustness question that was not addressed by the original $W=100$, $m=1$, $H=14$ design. Across $W\in\{100,200,400\}$, $m\in\{1,3,5\}$, and $H\in\{5,14,20\}$ at $\sigma=0.10$, all 108 channel--configuration combinations remain positively aligned with the imposed coupling, with $T\ge16.49$ and a positive realization-level slope confidence interval at every tested grid point (Table~\ref{tab:param_robustness}). The original configuration remains strongly responsive for all four hidden forms and was not replaced by a post-hoc optimum.

The three parameters play different roles (Fig.~\ref{fig:param_effects}). Window length has the strongest marginal effect: mean $T$ rises from 23.65 at $W=100$ to 36.30 at $W=200$ and 43.05 at $W=400$. This gain is expected in a stationary benchmark because larger windows provide more training examples and more held-out errors from which to estimate a distribution. It should not be interpreted as a universal recommendation for $W=400$. The revised intermittent benchmark in Figs.~\ref{fig:intermittent} and \ref{fig:n2_tracking_quality} makes the complementary cost explicit: increasing the rolling window from 200 to 400 raises the LightGBM Pearson tracking correlations with the known window-mean coupling from 0.860 to 0.911 for FEDE $\Delta$ and from 0.863 to 0.924 for FEDE early slope, but broadens the transitions around each switch. In a nonstationary application, the useful $W$ must therefore balance statistical stability against temporal resolution.

History length has a smaller systematic effect, with the highest mean $T$ at $m=1$ and modest reductions at $m=3$ and 5. This is consistent with the short-memory construction of the synthetic process and should not be generalized to empirical systems with longer memory. The maximum horizon is comparatively insensitive over the tested range: mean $T$ is 35.96, 33.40, and 33.65 for $H=5$, 14, and 20. Thus the main result does not depend critically on the particular choice $H=14$, while still requiring a genuinely multi-horizon profile: for the endpoint definition $\Delta\mathrm{FEDE}=\mathrm{FEDE}(H)-\mathrm{FEDE}(1)$, $H=1$ would be degenerate by construction.

\subsection{Scope, limitations, and future validation}
The study remains deliberately controlled. The synthetic benchmarks use one stochastic driver family, one target architecture, a finite set of hidden nonlinear mappings, one nonlinear prediction engine, and a basic CCM implementation without lag scanning, surrogate testing, or multivariate conditioning. The parameter study establishes robustness only over the tested ranges, not universal invariance. FEDE also remains model-relative: if the prediction engine cannot extract a real interaction, the entropy descriptor cannot recover information that is absent from its forecast errors. Conversely, a highly flexible predictor can exploit spurious structure unless the validation design, source controls, and temporal split are chosen carefully.

Real observational data add challenges that are intentionally absent here: common drivers, delayed feedback, changing regimes, missing observations, unequal sampling, multiple candidate sources, and limited sample sizes. These issues motivate the next validation stage. The broader literature shows that directional time-series questions arise across ecological interactions, epidemiological surveillance, hydrology and climate, physiological/neuroscience signals, and financial/economic systems \cite{sugihara2012,maher2016,bonotto2022,hasanzadeh2021,isufaj2025}. Future work should therefore test whether the synthetic advantages demonstrated here---functional-form robustness, source specificity, and multi-horizon sensitivity---survive on observational datasets where the ground truth is incomplete and competing explanations cannot be controlled by construction. Such studies should include richer comparators where appropriate, including lagged or surrogate-tested CCM/EDM, nonlinear or neural Granger variants, and formal forecast-loss tests \cite{tank2021,wismuller2021,grant2026}. An especially relevant extension is to replace the conventional residual-variance score in matched Granger-style models by residual-distribution entropy and test, on the same restricted/full residuals, whether the sensitivity observed for FEDE arises primarily from the distributional observable, from the multi-horizon construction, or from their combination.

The present evidence therefore supports a deliberately limited claim. The FEDE predictive-causality framework is not a replacement for Granger causality, CCM, or conventional forecast evaluation, and its use of the term predictive causality does not imply structural causal identification. Its contribution is a new \emph{horizon-resolved predictive observable}: by comparing the entropy evolution of matched signed forecast-error distributions, it converts multi-horizon forecasts into a directional descriptor that remains sensitive across noise, source controls, nonlinear functional forms, and a broad internal parameter grid. The main practical value demonstrated here is the combination of \emph{representation flexibility}, \emph{directional specificity}, and \emph{robust multi-horizon response} at little additional cost once nonlinear forecasts are already available.

\section{Conclusions}
This work introduced the forecast-error distribution entropy (FEDE) predictive-causality framework as a horizon-resolved approach to directional predictive contribution. Predictive causality is used here strictly in the Granger operational sense: the source past must provide additional out-of-sample predictive information beyond the target's own past, within the specified model and conditioning set. The construction compares matched target-only and target-plus-source forecasts and measures how the entropy of the signed forecast-error distribution evolves from short to longer lead times. The method is therefore distinct from correlation, one-step residual comparison, and state-space reconstruction: it uses the geometry of the forecast-error distribution across horizon as the measured object.

The controlled experiments establish five complementary properties. First, FEDE detects a hidden nonlinear predictive channel even when raw linear association is near zero, provided that the forecasting engine can extract the underlying information. Second, its response is source- and direction-specific: the true $Y\rightarrow X$ channel is separated from the reverse direction and from independent periodic, quasiperiodic, chaotic, and AR(1) auxiliaries. Third, FEDE remains positively aligned across four substantially different hidden nonlinear mappings without changing its representation, whereas the OOS polynomial Granger comparator is extremely strong only when its prescribed degree matches the true quadratic form. Fourth, in the intermittent benchmark the LightGBM FEDE response tracks the time-varying imposed coupling, with the $W=400$ early-slope statistic reaching Pearson $r_P=0.924$ against the known window-mean coupling and exposing the expected stability--temporal-resolution trade-off relative to $W=200$. Fifth, the positive stationary response survives all 108 tested combinations of window length, history length, and maximum forecast horizon, with $T\ge16.49$ throughout the predefined grid.

The comparative results also define where FEDE should and should not be used. A correctly specified low-dimensional parametric model can be both faster and more sensitive; in the matched $H_2$ case, degree-2 polynomial Granger reaches $T=74.21$ compared with FEDE's $T=29.00$. Basic CCM addresses a different state-space reconstruction question and should not be judged by the stochastic non-invertible benchmark alone. The total FEDE+LightGBM pipeline is computationally much more expensive than OLS-based Granger analysis, but once the multi-horizon LightGBM forecasts are available the entropy calculation adds only about 2.2\% to the shared forecast-core time. FEDE's advantage is therefore not universal accuracy or speed, but a comparatively uniform response when the interaction form is unknown and multi-horizon nonlinear forecasts are already part of the workflow.

The next step is validation on observational time series. Candidate domains include financial and economic systems, climate and hydrology, epidemiological surveillance, physiology and neuroscience, and ecological interaction networks, all of which have established traditions of predictive or state-space directional analysis \cite{sugihara2012,maher2016,bonotto2022,hasanzadeh2021,isufaj2025}. Such applications should test robustness to confounding, feedback, regime changes, missing data, multiple candidate sources, and uncertain time delays, and should compare FEDE with stronger domain-appropriate baselines rather than relying on the synthetic ranking reported here.

The main contribution of the study is consequently specific but generalizable: the FEDE predictive-causality framework turns the \emph{horizon evolution of forecast-error distribution entropy} into a model-relative directional predictive descriptor. Across the controlled tests performed here, this observable adds information beyond average forecast error, preserves specificity to the true source and direction, and remains stable across markedly different nonlinear forms and analysis scales. These properties establish FEDE as a sensitive and robust complementary tool for studying nonlinear predictive coupling when the interaction form is not known in advance, while preserving a clear distinction between predictive and structural causality.

\section*{Funding}
This research was supported by the Russian Science Foundation (grant no.~22-11-00055-P).

\section*{Code and data availability}
A persistent Zenodo record for the FEDE predictive-causality software and benchmark package is publicly available at DOI: \href{https://doi.org/10.5281/zenodo.23330544}{10.5281/zenodo.23330544}. The software package includes the reproducible Python implementation, benchmark scripts, configuration files, reference results, and a researcher-facing application workflow. It also contains numerical result tables underlying the principal method-comparison, parameter-robustness, and N2 time-resolved tracking analyses, together with the N2 settings used to generate the corresponding figures. The synthetic benchmark definitions, parameter settings, and analysis procedures required to interpret the results are described in the manuscript and accompanying repository.

\appendix
\section{Supplementary nonlinear diagnostics}\label{sec:appendix_n2}
\begin{figure}[H]
\centering
\includegraphics[width=\linewidth]{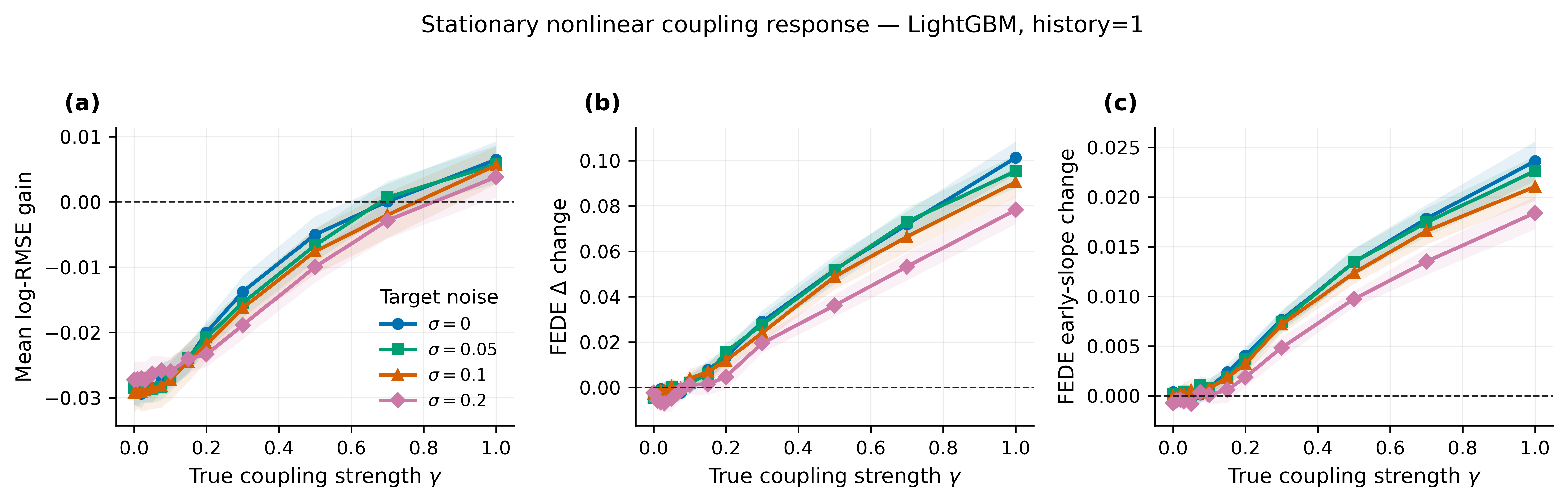}
\caption{Stationary hidden nonlinear coupling response for LightGBM across the complete noise grid. Bands show 95\% confidence intervals of the mean across 20 realizations.}
\label{fig:supp_nonlinear_noise}
\end{figure}

\begin{figure}[H]
\centering
\includegraphics[width=0.98\linewidth]{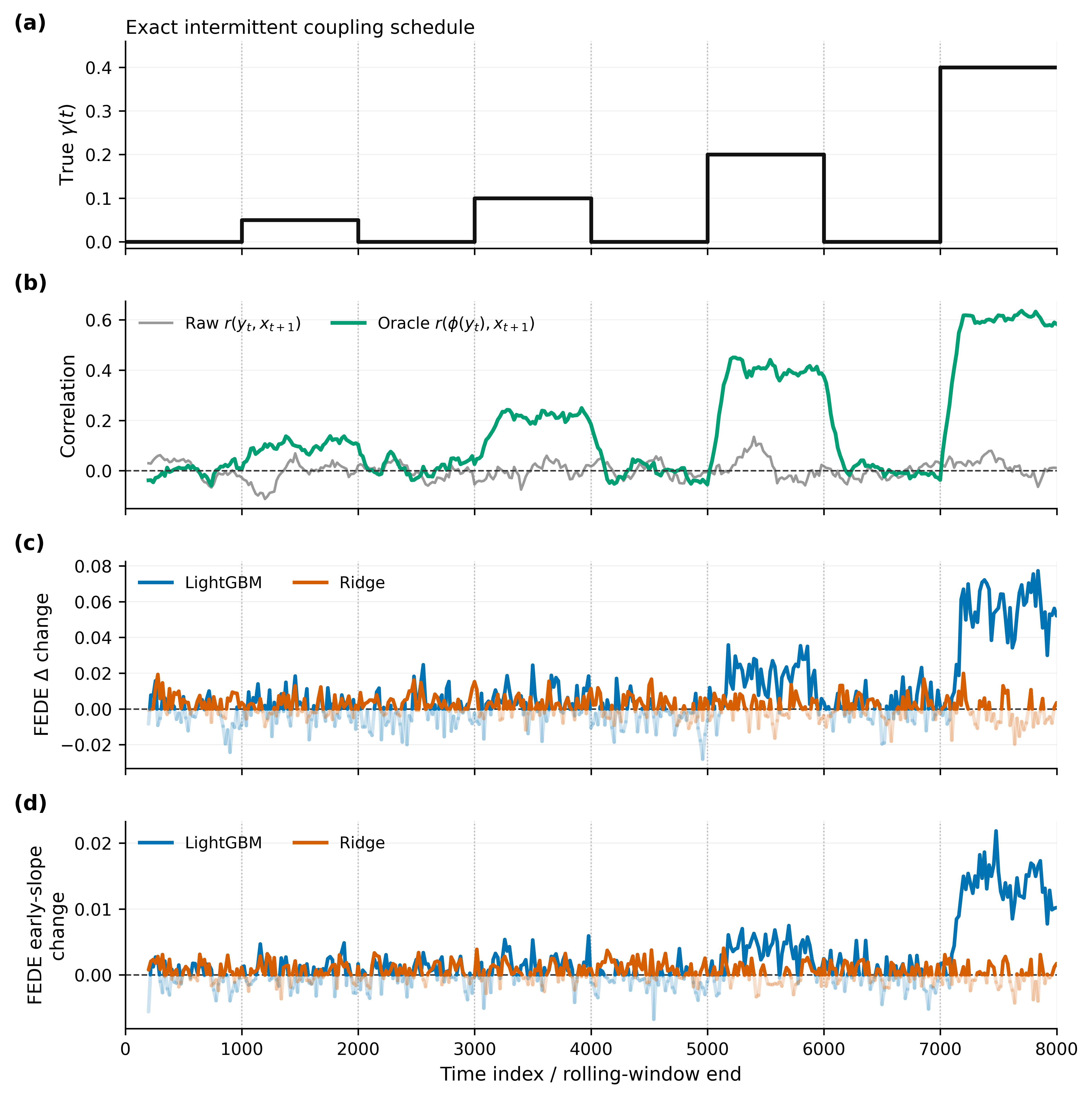}
\caption{Benchmark N2 supplementary time-domain comparison at $W=200$, $\sigma=0.10$, history $m=1$, rolling step 20, and $H=14$. The imposed schedule and all stochastic realizations are the same as in the $W=400$ main-text analysis. The shorter window preserves sharper temporal localization around switching boundaries but yields a visibly less stable FEDE response. Negative FEDE values are retained and displayed with lower opacity rather than clipped.}
\label{fig:supp_n2_w200}
\end{figure}

Figure~\ref{fig:supp_nonlinear_noise} shows the complete LightGBM stationary response across all four nonlinear noise levels. Figure~\ref{fig:supp_n2_w200} provides the $W=200$ time-domain comparison for benchmark N2 and complements the $W=400$ main-text visualization.

\section{Supplementary calibration-range diagnostic}
\begin{figure}[H]
\centering
\includegraphics[width=\linewidth]{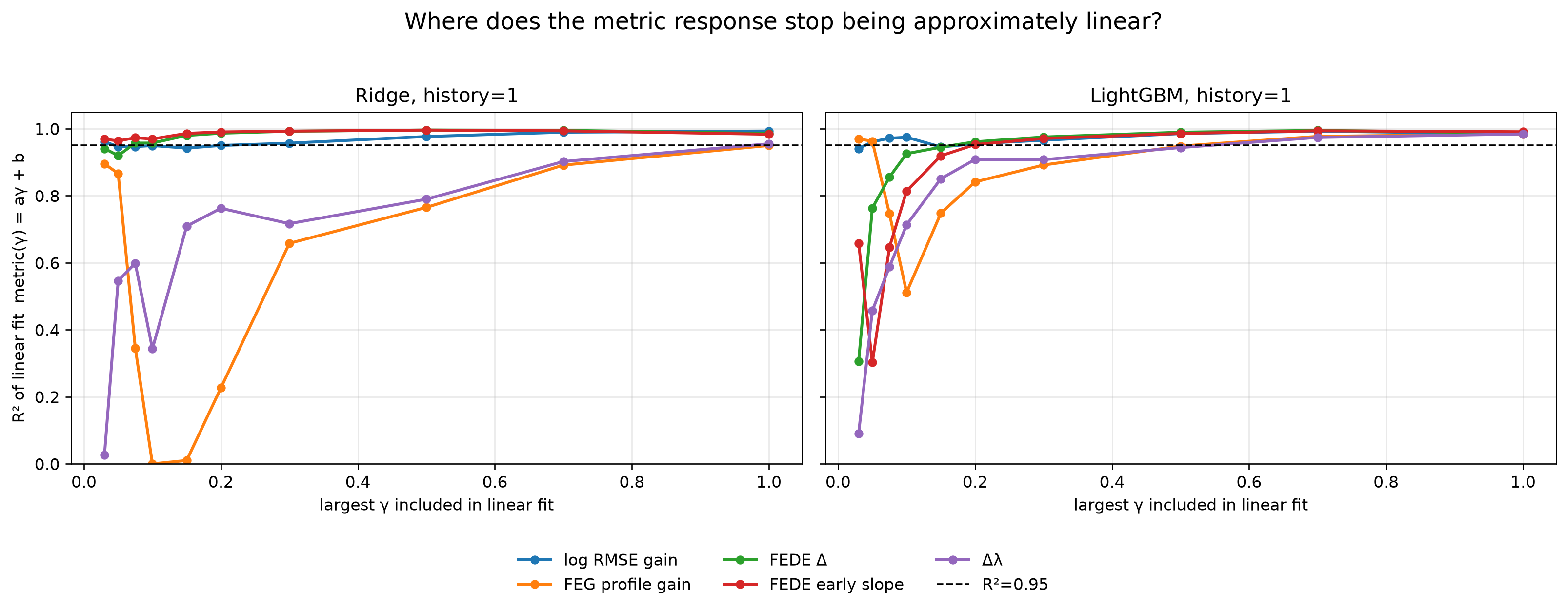}
\caption{Linearity diagnostic as the fitted coupling range is expanded. The main text therefore emphasizes the complete controlled calibration and the noise-dependent linearity analysis rather than any single very narrow range.}
\label{fig:breakdown}
\end{figure}

Figure~\ref{fig:breakdown} shows how the linear-fit $R^2$ changes as progressively larger maximum coupling values are included. It is retained as a diagnostic because very narrow low-$\gamma$ ranges can produce unstable $R^2$ values when the response amplitude is comparable to realization variability.

\end{document}